\documentclass[a4paper,11pt]{article}
\usepackage{jheppub}
\usepackage[utf8]{inputenc}
\usepackage{xspace}
\usepackage{cleveref}

\numberwithin{equation}{section}
\usepackage{booktabs}
\usepackage[dvipsnames]{xcolor}

\newcommand{\SU}[0]{\ensuremath{SU(3)_F}\xspace}
\newcommand{\EOS}{\texttt{EOS}\xspace}
\newcommand{\Bs}[0]{\ensuremath{B_s}\xspace}
\newcommand{\Bd}[0]{\ensuremath{B_d}\xspace}
\newcommand{\Bu}[0]{\ensuremath{B^+}\xspace}
\newcommand{\Bc}[0]{\ensuremath{B_c}\xspace}
\newcommand{\Bsb}[0]{\ensuremath{\bar B_s}\xspace}
\newcommand{\Bdb}[0]{\ensuremath{\bar B_d}\xspace}

\makeatletter
\g@addto@macro\bfseries{\boldmath}
\let\Hy@backout\@gobble
\makeatother

\title{Extracting production fractions of $b$ hadrons from exclusive semi-leptonic decays}
\preprint{Nikhef-2026-007, EOS-2026-02}

\author[a]{Carolina Bolognani,}
\emailAdd{carolina.bolognani@cern.ch}
\author[b]{Martin Jung,}
\emailAdd{martin.jung@unito.it}
\author[c]{Méril Reboud,}
\emailAdd{meril.reboud@cnrs.fr}
\author[d]{K. Keri Vos}
\emailAdd{k.vos@maastrichtuniversity.nl}

\affiliation[a]{Fakultät Physik, Technische Universität Dortmund, Otto-Hahn-Str.4, D-44221 Dortmund, Germany}
\affiliation[b]{Dipartimento di Fisica, Università di Torino \& INFN, Sezione di Torino, Via Pietro Giuria 1, I-10125 Turin, Italy}
\affiliation[c]{Université Paris-Saclay, CNRS/IN2P3, IJCLab, 91405 Orsay, France}
\affiliation[d]{Gravitational Waves and Fundamental Physics (GWFP), Maastricht University, Duboisdomein 30, NL-6229 GT Maastricht, the Netherlands}

\abstract{
Ratios of production fractions of $b$ hadrons are a dominant source of uncertainty in many LHC analyses, in particular in measurements with $B_s$ mesons.
The currently used value for $f_s/f_d$ is based on a combination of hadronic and semi-inclusive semi-leptonic decays, and relies in part on assumptions about the underlying decays that are hard to quantify.
We propose an independent alternative method to obtain this quantity by measuring ratios of the exclusive semi-leptonic decays $\bar B_{(s)} \to D_{(s)}^{(*)} \ell \bar\nu$.
This method benefits from significant cancellations of both experimental and theoretical uncertainties, as well as robustness against potential contamination from heavy physics beyond the Standard Model.
As a proof of principle, we show that current measurements constrain $f_s/f_d$ with an uncertainty of $7\%$, dominated by present experimental uncertainties.
This method can also be applied to other ratios of production fractions involving heavier $b$ hadrons, such as $B_c$ or $\Lambda_b$.
}

\begin{document}
\maketitle

\pagenumbering{arabic}
\section{Introduction}

Weak decays of \Bs mesons provide stringent tests of the Standard Model (SM) of particle physics as well as physics beyond it (BSM), complementary to their \Bd, \Bu counterparts. 
While a number of results are available from the $B$ factories, the majority of \Bs results come from experiments at the Large Hadron Collider (LHC), a situation that is unlikely to change in the near future.
A challenge at hadron colliders is to obtain absolutely normalised branching fractions, as no precise determinations of the production cross sections for the different $b$-hadron species are available.
Therefore, ratios of branching fractions are generally measured and are then multiplied by the (absolutely normalised) branching fractions of the corresponding normalisation modes.
Since no sufficiently precise result exists for any \Bs branching fraction, \Bd and \Bu branching fractions are usually used for that purpose, as they can be obtained at the $B$ factories, where the corresponding production fractions have been measured at the per cent level \cite{HeavyFlavorAveragingGroupHFLAV:2024ctg,Bernlochner:2023bad,Jung:2026ewj}.
The remaining difficulty in extracting $B_s$ branching fractions is then to obtain the required ratios of production fractions in the corresponding experimental environment.

These production fractions, $f_q \equiv \mathcal{B}(b\to \bar B_q)$, and their kinematic dependencies have been studied in detail in proton-proton ($pp$), $e^+e^-$ and $p\bar{p}$ collisions \cite{HeavyFlavorAveragingGroupHFLAV:2024ctg}.
The most recent determination by the LHCb collaboration combines five analyses using hadronic and semi-inclusive semi-leptonic decays, quoting a $3\%$ uncertainty on $f_s/f_d$ assuming $f_d=f_u$ \cite{LHCb:2021qbv}.
While this is an achievement, it is not sufficient for future precision measurements, many of which will be limited by the resulting systematic uncertainty.
In the key measurement of $B_s\to \mu^+\mu^-$, for example, the limited knowledge of the production fractions yields already the largest systematic uncertainty, both for CMS \cite{CMS:2022mgd} and LHCb \cite{LHCb:2021awg}.
A significant improvement in the precision of $f_s/f_d$ is therefore crucial to fully exploit the existing and expected LHC datasets.

The general idea behind obtaining $f_s/f_d$ from experimental data is straightforward: it requires a precise measurement of a \Bs over \Bd decay, paired with an ideally similarly precise theoretical determination of this ratio.
These two requirements do not necessarily align, however.
For instance, experimentally, a hadronic decay with a large branching fraction into only charged particles would be ideal to achieve a high reconstruction efficiency, for instance $B_s\to J/\psi(\to \mu^+\mu^-) \phi(\to K^+K^-)$.
Theoretically, however, the precision of calculations of hadronic decays is generally limited; even for modes that are more amenable to a direct calculation than $B_s\to J/\psi \phi$, improvements beyond $\Lambda/m_b \sim 10\%$ or $m_s/\Lambda \sim 30\%$ are extremely challenging.

Alternatively, semi-leptonic decays can be used, which are generally under better theoretical control, though they pose an experimental challenge due to the missing neutrino in the final state.
Existing analyses have focused on semi-inclusive modes $\bar B_{(s)}\to D_{(s)}X\ell\bar\nu$ to probe the combination $f_s/(f_u+f_d)$ by relating them to the corresponding fully inclusive modes, 
whose ratio can be calculated in a well-defined expansion in $1/m_b$ and $\alpha_s$.
This method, however, requires assumptions whose precision is difficult to quantify. 

We, therefore, propose to instead determine $f_s/f_d$ from measurements of the ratios of \emph{exclusive} semi-leptonic decay rates, such as
\begin{equation}\label{eq:ratdef}
   R_{sd}^{D} \equiv \frac{\Gamma(\Bsb\to D_s \ell \bar\nu)}{\Gamma(\Bdb\to D_d \ell \bar\nu)} \qquad \mbox{and}\qquad R_{sd}^{D^*} \equiv \frac{\Gamma(\Bsb\to D_s^{*} \ell \bar\nu)}{\Gamma(\Bdb\to D_d^{*} \ell\bar\nu)} \,.
\end{equation}
In this equation $\ell$ stands for any charged lepton, but in the following we only keep $\ell = \mu$, which is experimentally the most accessible.
These decays have several advantages:
The CKM matrix element $|V_{cb}|$ cancels in these ratios, as do part of the form factor uncertainties, where their correlations are known.
The modes are related by \SU, and symmetry-breaking contributions are additionally suppressed by heavy-quark symmetry \cite{Isgur:1989vq, Isgur:1990yhj,Kobach:2019kfb}.
For that reason, these ratios are only very weakly dependent on BSM contributions.
Most importantly, however, they only depend on local $\bar B\to D^{(*)}$ and $\bar B_s\to D_s^{(*)}$ form factors, which are already known to good precision and are expected to be improved in the future using Lattice QCD.
On the experimental side, the similarities between the two modes also help reduce systematic uncertainties.
As such, the proposed approach is systematically improvable.
We expect these advantages to enable a precise determination of $f_s/f_d$ with quantifiable theoretical uncertainties.

This paper is organised as follows:
In \cref{sec:currentfsfd}, we review the currently available determinations of $f_s/f_d$.
In \cref{sec:theory}, we discuss the theory predictions of the ratios in \cref{eq:ratdef}.
In \cref{sec:method}, we discuss how these ratios can be used to extract $f_s/f_d$ and, as a proof of principle, derive it from existing measurements.
In \cref{sec:NPpollution}, we show explicitly that possible heavy NP contributions have a very minor effect on the experimental ratios. 
Finally, in \cref{sec:fcfd}, we discuss the generalisation of our method to other modes, in particular for the determination of $f_c/f_d$.

\section{Current status of $B_s$ fragmentation functions} \label{sec:currentfsfd}
The way ratios of production fractions $f_a/f_b$ are generally extracted at hadron colliders is to assume knowledge of a ratio of decay rates of the corresponding $b$ hadrons $H_i$, \emph{i.e.}, using the efficiency-corrected yields
\begin{equation}
    \frac{n_{\rm corr}(H_a\to X_a)}{n_{\rm corr}(H_b\to X_b)} = \frac{\Gamma(H_a\to X_a)}{\Gamma(H_b\to X_b)} \frac{f_a}{f_b} \frac{\tau_{H_a}}{\tau_{H_b}} \,.
\end{equation}
To extract the ratios $f_s/f_{d,u}$ (\emph{i.e.}, $H=B,a=s,b=d,u$ in the above equation), ratios of hadron decays that form \SU-symmetry partners can be chosen, although this is not a strict requirement.
Since the generic uncertainty from \SU-symmetry breaking effects is generally insufficient to obtain precise fragmentation ratios, the modes are chosen to either have an additional suppression of the breaking or for the breaking to be to some extent calculable.

The most recent determinations of $f_s/f_d$ by the LHCb collaboration read \cite{LHCb:2021qbv}
\begin{align}\label{eq:fsfd}
    f_s/f_d \, (7 \; \rm{TeV}) &  = 0.2390 \pm 0.0075 \ , \nonumber\\
    f_s/f_d \, (8 \; \rm{TeV}) &  = 0.2385 \pm 0.0075 \ , \nonumber \\
    f_s/f_d \, (13 \; \rm{TeV})&  = 0.2539 \pm 0.0079 \ ,
\end{align}
where $f_d=f_u$ is assumed.
These values are obtained by combining five LHCb analyses \cite{LHCb:2021qbv}:
\begin{itemize}
    \item Hadronic $\Bs\to D_s^-\pi^+$ and $\Bd\to D^- K^+$ decays \cite{LHCb:2013vfg, LHCb:2020zae} for the absolute determination;\footnote{The ratio $B_s^0\to D_s^-\pi^+$ to $B_d^0\to D^- \pi^+$ was also used to determine the relative $f_s/f_d$ ratio at $7 \; \rm{TeV}$ \cite{LHCb:2013vfg}}
    \item Semi-inclusive semi-leptonic $\bar{B}_{(s)}\to D_{(s)}X\mu\bar\nu$ decays \cite{LHCb:2011leg, LHCb:2019fns} for the absolute determination;
    \item Charmonium $\Bs\to J/\psi \phi$ and $B^+\to J/\psi K^+$ decays to determine the $p_T$ dependence \cite{LHCb:2019lsv}. 
\end{itemize}
The dependence of $f_s/f_d$ on the pseudorapidity was also studied, but no such dependence was observed in the measurements considered in Ref.~\cite{LHCb:2021qbv}.
Below, we briefly discuss the hadronic and semi-leptonic determinations.
First, however, we comment on the approximation $f_u/f_d = 1$.
Presently, it is common practice to make this assumption in analyses at hadron colliders.
While this is likely to be a reasonable approximation, given that most $b$-hadrons are produced via $b\bar b$, it is also true that the initial state $pp$ does not correspond to an isospin singlet and can therefore lead to a violation of that expectation.
The extent to which this affects the ratio $f_u/f_d$ is ultimately an experimental question and will, in particular, depend on transverse momentum.
LHCb results for $B\to D\bar D$ decays \cite{LHCb:2013sad} hint at a significant deviation from unity \cite{Jung:2014jfa,Davies:2023arm}, while the two dedicated analyses by the CMS collaboration \cite{CMS:2022wkk,CMS:2026kkx} find results consistent with unity at higher values of $p_T$. 
In light of this, $f_u$ should be clearly distinguished from $f_d$ and an additional uncertainty assigned to the assumption $f_u/f_d = 1$.
A dedicated analysis in the LHCb environment would be very welcome to clarify this issue.

\subsection{Determination from $B_s\to D_s^- \pi^+$ and $B_d\to D^- K^+$ }
The idea behind obtaining $f_s/f_d$ from the hadronic decays $B_s\to D_s^-\pi^+$ and $B_d\to D^-K^+$ (see, e.g., Ref.~\cite{Fleischer:2010ay}) is that these decay modes are pure tree decays which factorise in the heavy-quark limit \cite{Beneke:2000ry}.
The radiative corrections to these modes have been calculated to $\mathcal O(\alpha_s^2)$ \cite{Huber:2016xod}, while the required $B_s\to D_s$ and $B_d\to D$ form factors are the same as those studied in this paper. While power corrections of $\mathcal O(1/m_c)$ are absent outside these form factors \cite{Beneke:2000ry,Bordone:2020gao}, the size of power corrections
of $\mathcal{O}(\Lambda/m_b)$ is under discussion.
Under the assumption that they are small -- as estimated in Ref.~\cite{Bordone:2020gao} -- the SM prediction of the ratio of branching fractions has reached the few-per-cent level, which in principle allows a determination of $f_s/f_d$ at the same level of precision.
However, it has recently been pointed out that, although the ratio of branching fractions obtained from factorisation is consistent with experimental measurements, the individual amplitudes extracted using the semi-inclusive $f_s/f_d$ value differ by around $20\%$ from their SM predictions \cite{Bordone:2020gao}.
This tension has not been understood so far, questioning the theoretical basis for this method of extracting $f_s/f_d$.
Missing power corrections may be universal for the $B_s$ and $B_d$ modes, and hence cancel in the ratio, but this is not guaranteed.
As a result, the uncertainty in the SM prediction of the ratio in question may be much larger than previously expected, invalidating prior error estimates and casting doubt on the competitiveness of this approach.

\subsection{Inclusive Semi-leptonic $\bar B_{(s)}\to D_{(s)} X \ell \nu$ Decays}
Another approach to obtain $f_s/f_d$ would be to use fully inclusive $\bar B_{(s)}\to X_c \ell\bar \nu$ modes, which is one of the ways the production fractions at the $B$ factories have been determined \cite{Belle:2002lms}.
Such fully inclusive rates are independent of the spectator quark flavour, up to corrections of order $1/m_b^2$ and higher (see e.g.~\cite{Mannel:2023yqf, Finauri:2025ost}),
\begin{equation}\label{eq::BsdXc}
\frac{\Gamma(\bar B_s\to X_c\ell\bar \nu)}{\Gamma(\bar B_d\to X_c\ell\bar \nu)}\approx 1+\mathcal O(1/m_b^2)\,.
\end{equation}

However, measuring fully inclusive modes directly is currently not feasible at hadron colliders.
Instead, (quasi-)inclusive modes $\bar B_{(s)}\to D_{(s)} X \ell\bar \nu$ are used.
The idea is that the majority of charmed states produced in a $b\to c\ell\bar\nu$ transition of a $B$ meson decays further into a final state including a $D_{(s)}$ meson,\footnote{%
    In practice, ratios with respect to the sum of $B_d^0$ and $B^+$ modes are measured and $f_s/(f_u+f_d)$ is extracted, from which $f_s/f_d$ can then be obtained assuming $f_u = f_d$.
} \emph{i.e.}, 
\begin{equation}\label{eq::BDXapprox}
    \Gamma(\bar B_{(s)}\to D_{(s)}X\ell\bar \nu) = \Gamma(\bar B_{(s)}\to X_c\ell\bar\nu)-\delta_{(s)}\,.
\end{equation} 
To the extent that $\delta_{(s)}$ is very small or correctly modelled, the properties of the fully inclusive decay can be applied.
This is one of the key challenges of this approach, which we discuss in more detail below.
At hadron colliders, an additional complication arises because the initial state can involve any $b$-flavoured hadron.
Such cross-feed contributions to the same final state from other mesons and higher excitations need to be corrected for in the experimental analysis, which relies on models used in simulations and symmetry assumptions. 

On the theoretical side, obtaining $f_s/f_d$ from these decays requires knowledge of the power corrections in~\cref{eq::BsdXc}.
For the $B_d$ system, these corrections are known with good precision from moments of the inclusive $B_d\to X_c \ell \nu$ spectra \cite{Finauri:2023kte,Bernlochner:2022ucr, Carvunis:2025vab}. 
However, for the $B_s$ system, these corrections are not known as experimental data is not available.\footnote{%
    Another approach would be to determine the inclusive rates directly using Lattice QCD, as discussed in Refs.~\cite{Gambino:2022dvu,Gambino:2023xoe,DeSantis:2025qbb,DeSantis:2025yfm}
} 
Therefore, the corrections to~\cref{eq::BsdXc} can only be assessed by relating the $B_d$ and $B_s$ HQE elements using \SU symmetry and estimating the \SU-breaking effects. The most recent estimate of these corrections is \cite{Bordone:2022qez,Bigi:2011gf}
\begin{equation}\label{eq::BsXcest}
    \left|\frac{\Gamma(\bar B_s\to X_c\ell\bar \nu)}{\Gamma(\bar B_d\to X_c\ell\bar \nu)}\right|_{\mathrm{th}}\! - 1 = -(1.8\pm 0.8)\% \ .  
\end{equation}
We note that both the size and the uncertainty of this estimate have increased compared to the earlier estimate in Ref.~\cite{Bigi:2011gf}, owing to sizeable contributions at $1/m_b^3$ and to an estimate of higher-order contributions.
It would be interesting to confront this estimate with experimental data to further test the \SU assumption.
Nevertheless, combined with precise experimental data, these fully inclusive decays could thus serve as a means to extract $f_s/f_d$.
However, this estimate only applies to fully inclusive decays, as the theory for semi-inclusive decays does not yet exist.

The more difficult point in the extraction of $f_s/f_d$ from these decays is the reliability of the size and composition of $\delta_{(s)}$ in \cref{eq::BDXapprox}, for which a combination of data, models and approximations is used.
There are two classes of contributions to $\bar B_{(s)}\to X_c\ell\bar \nu$ that are not covered by $\bar B_{(s)}\to D_{(s)}X\ell\bar\nu$, involving different $D$ mesons and charmed baryons.
The latter are assumed to be negligible.
The former are assumed to be saturated by $B\to D_s^{(*)}K\ell\nu$ for $\delta$. For $\delta_s$, a sizeable correction comes from $B_s\to DKX\ell\nu$ due to above-threshold $D_s$ resonances.
Both are modelled and considered part of the corrected $B_{(s)}\to D_{(s)}X\ell\nu$ samples. 

The difficulty with neglecting or correcting for the missing contributions is that it requires at least approximate knowledge about their size.
While in some cases the necessary explicit experimental data exists, others have to be estimated.
These estimates are also challenging: it is known that the sum of exclusive modes does not saturate the measured inclusive rate (see, for instance, Refs.~\cite{Bernlochner:2012bc,Rudolph:2018rzl,Herren:2026pni}) even when including estimates for the unmeasured modes.
This indicates either that the existing estimates for the missing modes are inadequate or that unknown systematic effects affect these measurements (see, for instance, Ref.~\cite{Herren:2026pbh}).

The experimentally available information on $B\to DX\ell\nu$ decays is scarce \cite{CLEO:1990fzo,BaBar:2007xlq,LHCb:2018azb}, see also Ref.~\cite{LHCb:2019fns}. Performing a combined fit to these results under the assumption of isospin for the involved rates, including the semi-inclusive rate for $B^{0,+}\to \bar D X\ell\nu$, where $\bar D$ includes both neutral and charged $D$ mesons,\footnote{Note that isospin is \emph{not} a good symmetry for the individual $D$-meson final states, mostly due to threshold effects in $D^*\to D\pi$.} we obtain a good fit ($\chi^2/\text{dof}=2.5/3$), and in particular
\begin{equation}\label{eq::DXoXc}
    \frac{\Gamma(B\to \bar DX\ell\nu)}{\Gamma(B\to X_c\ell\nu)} = 0.90\pm0.05\,,
\end{equation}
which is in line with the individual results listed in Ref.~\cite{ParticleDataGroup:2024cfk}. 
This result might point towards a sizeable deviation from the assumption that $\delta_{(s)}$ in \cref{eq::BDXapprox} is small, but is not yet sufficiently precise to establish that conclusion, emphasising yet again the importance of obtaining additional experimental information on semi-inclusive $B\to DX\ell\nu$ decays. 
We stress that in \cref{eq::DXoXc} only measurements of $B$ mesons with light spectator quarks enter. This differs from the measurement used to extract $f_s/f_d$, where, as noted above, a priori all $b$-hadron species contribute.
Despite ongoing lattice calculations, future theory improvements are particularly challenging as the theory for the semi-inclusive decay does not exist.
In the $B_s$ case, the corresponding ratio has been measured to be \cite{Belle:2015ftp,Belle:2012mwf}
\begin{equation}\label{eq::DsXoXc}
    \frac{\Gamma(B_s\to \bar D_sX\ell\nu)}{\Gamma(B_s\to X_c\ell\nu)} = 0.86\pm0.17 \quad [0.82\pm0.15]\,,
\end{equation}
where the second number uses the estimate in \cref{eq::BsXcest} instead of the corresponding measurement \cite{Belle:2012mwf}. As explained, this value is expected to lie below unity, given the sizable $B_s\to DX\ell\nu$ fraction.
This has to be corrected for if \cref{eq::BDXapprox} is to be used, which is difficult, however, since the $\Bs\to \bar DX\ell\nu$ fraction is very uncertain. Again, the rather low central value is not conclusive, given the large uncertainties.

\section{Theory predictions for exclusive $B\to D^{(*)}$ and $B_s\to D_s^{(*)}$} \label{sec:theory}

The semi-leptonic rates for $\bar B_{(s)}\to D_{(s)}^{(*)}\ell\bar \nu$ are expressed in a standard way by calculable coefficients times hadronic form factors that parametrise the non-perturbative matrix elements, up to electromagnetic corrections that are much smaller than the present uncertainties \cite{Beneke:2021jhp}. 
The dominant sources of theory uncertainty 
in the calculation of the rates are therefore
$V_{cb}$, which cancels in $R_{sd}^{D^{(*)}}$, and the hadronic form factors.

\subsection{Hadronic form factors}
In the limit of equal light-quark masses and negligible electromagnetic corrections, \SU symmetry relates hadronic matrix elements between initial and final states that belong to the same multiplet of the symmetry.
In the simplest cases, as the one at hand, these relations amount directly to equalities between matrix elements for, \emph{e.g.}, $B_d$ and $B_s$ decays and consequently their form factors.
Generically, the leading corrections to these limits scale with the light-quark-mass differences like $(m_q-m_{q^\prime})/\Lambda_\mathrm{QCD}\sim m_s/\Lambda_\mathrm{QCD}\sim30\%$.
Due to heavy-quark symmetry, the symmetry-breaking effects in $\bar B\to D^{(*)}\ell\bar \nu$ transitions are further suppressed at zero-recoil. The zero-recoil point 
provides a normalisation of the matrix element, which is independent of the spectator quark up to $1/m_c^2$ corrections \cite{Isgur:1989vq, Isgur:1990yhj}. 
A global analysis of the available $\bar B_{(s)}\to D_{(s)}^{(*)}$ form factor information based on the heavy-quark expansion has recently been performed in Ref.~\cite{Bordone:2025jur}, confirming this picture, see also Ref.~\cite{Bernlochner:2022ywh}. 

The kinematic dependence of the form factors can be expanded in the conformal variable $z$, defined below, with $|z|_\mathrm{max} \sim \mathcal O(\%)$ and coefficients limited by unitarity \cite{Boyd:1997kz}.
On the other hand, corrections to the heavy-quark limit scale like $\Lambda_\mathrm{QCD}/m_{c,b}$.
The combined $z$ and heavy-quark expansion is therefore expected to show \SU-breaking effects only at the level of a few per cent \cite{Kobach:2019kfb}. \\

Going beyond this generic expectation, we can choose to use the explicit calculations available for the corresponding form factors in order to obtain the ratios $R_{sd}^{D^{(*)}}$.
To that aim, we adopt in the following the parameterisation of Ref.~\cite{Boyd:1997kz} and describe the $M_1\to M_2$ form factors with the variable
\begin{equation}
    z(q^2) = \frac{\sqrt{s_+ - q^2} - \sqrt{s_+ - s_0}}{\sqrt{s_+ - q^2} + \sqrt{s_+ - s_0}} \,,
\end{equation}
where $s_\pm = \left(m_{M_1} \pm m_{M_2}\right)^2$ and we used the optimal value $s_0 = s_+ \left( 1 - \sqrt{1 - \frac{s_-}{s_+}}\right)$.
The form factors are parametrised as~\cite{Boyd:1997kz}
\begin{equation}
    f(q^2) = \frac{1}{P(q^2) \phi(q^2)} \sum_{n=0}^N a_n^{(f)} z(q^2)^n \,,
\end{equation}
where $P(q^2)$ is a so-called Blaschke factor that implements the known poles of the form factor below $s_+$, and $\phi(q^2)$ is a known outer function that renders the unitarity bounds diagonal in the expansion parameters.
Neglecting the branch cuts below the pair production threshold,\footnote{%
    The size of these branch cuts has been estimated, for example, in Refs.~\cite{Boyd:1995sq,Caprini:1995wq}.
    Accounting for them in a model-independent way requires either changing the expansion polynomials as suggested in Ref.~\cite{Gubernari:2023puw}, or working with a non-diagonal bound as done, e.g., in Ref.~\cite{Flynn:2023qmi}.
    See also the discussions in Ref.~\cite{Gopal:2024mgb}.
} the unitarity bounds take the form
\begin{equation}
    \chi_\Gamma = \sum_f \sum_{n=0}^N {a_n^{(f)}}^2 < 1 \,,
\end{equation}
where the sum runs over all the form factors $f$ with the same operator structure $\Gamma$.
In the present analysis, these bounds are not imposed in the likelihood, and we only check their saturation \textit{a posteriori}.
The contribution of the form factors to the different bounds is summarised in \cref{tab:dipersive_bounds}.

\begin{table}[t]
    \centering
    \begin{tabular}{lccc}
    \toprule
        Bound       & effective susceptibility & form factors & Bound state masses [GeV] \\ 
    \midrule
        $0^+_V$     & $6.204\cdot10^{-3}$ & $f_0^{P\to P}$ & 6.704, 7.122 \\
        $0^-_A$     & $1.942\cdot10^{-2}$ & $\mathcal{F}_2^{P\to V}$ & 6.275, 6.871, 7.250 \\
        $1^-_V$     & $5.131\cdot10^{-4}\,\mathrm{GeV}^{-2}$ & $f_+^{P\to P}, g^{P\to V}$ & 6.329, 6.910, 7.020 \\
        $1^+_A$     & $3.894\cdot10^{-4}\,\mathrm{GeV}^{-2}$ & $\mathcal{F}_1^{P\to V}, f^{P\to V}$ & 6.739, 6.750, 7.145, 7.150 \\
        $1^-_T$     & $4.898\cdot10^{-4}\,\mathrm{GeV}^{-2}$ & $f_T^{P\to P}, T_1^{P\to V}$ & 6.329, 6.910, 7.020 \\
        $1^+_{T5}$  & $2.715\cdot10^{-4}\,\mathrm{GeV}^{-2}$ & $T_2^{P\to V}, T_{23}^{P\to V}$ & 6.739, 6.750, 7.145, 7.150 \\
    \bottomrule
    \end{tabular}
    \caption{Summary of the contributions to the dispersive bounds and the values used for the corresponding susceptibilities taken from Refs.~\cite{Bigi:2017jbd,Harrison:2024iad}.
    The effective susceptibilities are obtained by subtracting the 1-particle contributions to the susceptibilities.
    $P\to P$ includes $B\to D$ and $B_s\to D_s$ transitions and $P\to V$ includes $B\to D^*, B_s\to D_s^*$ and $B_c\to J/\psi$ transitions.
    The $\bar{b}c$ bound states masses are extracted from Refs.~\cite{Godfrey:2004ya,Dowdall:2012ab,CMS:2019uhm,LHCb:2019bem} and are not varied in our analysis.}
    \label{tab:dipersive_bounds}
\end{table}

We constrain the so-called BGL parameters $a_i^{(f)}$ to a combination of light-cone sum rules and lattice QCD estimates, which we detail in \cref{app:form-factor-constraints}.
This analysis is performed using the publicly available software \EOS~\cite{EOSAuthors:2021xpv} version 1.0.20~\cite{EOS:v1.0.20}.
Posterior samples are drawn using the nested sampling algorithm~\cite{Higson:2018} as implemented in the \texttt{dynesty} software~\cite{Speagle:2020,dynesty:v3.0.0}.
The codes used to run our analysis and all our results are available in the analysis repository~\cite{EOS-DATA-2026-02}.

We obtain a good fit already with a truncation order $N = 2$.
A goodness-of-fit summary is provided in \cref{tab:ff_gof}.
Since the LQCD inputs are obtained from fits that use equivalent parametrisations as the one used in the present analysis,
the $\chi^2$ values should not be interpreted in a statistical manner.
The excellent global fit quality, however, indicates that the theoretical inputs are compatible and that we can proceed with predictions, see also the discussions in Refs.~\cite{Martinelli:2023fwm,  Bordone:2024weh, Bernlochner:2022ywh, Fedele:2023ewe, Ray:2023xjn}.

\begin{table}[ht]
    \centering
    \begin{tabular}{lcccc}
    \toprule
       constraint           & number of constraints & number of parameters & $\chi^2$ & p-value [\%]\\
    \midrule
       $B\to D$ LQCD        & 12 & 5  & 8.5  & \\
       $B\to D^*$ LQCD      & 24 & 17 & 18.9 & \\
       $f_0^{B_{(s)}\to D_{(s)}}$ LQCD & 1 & 6 & 1.1 & \\
       $B_{(s)}\to D_{(s)}^*$ LQCD & 48 & 34 & 13.4 & \\
       $B_s\to D_s$ LQCD    & 5 & 5 & 0.6 & \\
    \midrule
       $B\to D^{(*)}$ LCSR  & 33 & 22 & 13.2 & \\
       $B_s\to D_s^{(*)}$ LCSR  & 33 & 22 & 5.6 & \\
    \midrule
       all & 156 & 44 & 61.3 & 100.00 \\
    \bottomrule
    \end{tabular}
    \caption{Goodness-of-fit summary for a truncation order $N = 2$.
    The constraints used in this analysis are listed in \cref{app:form-factor-constraints}.
    As explained in the main text, the p-values denoted in this table are only indicative since the corresponding inputs already stem from a fit.}
    \label{tab:ff_gof}
\end{table}

\subsection{Predictions for exclusive ratios}
We can now predict the ratio $R_{sd}^D$ and $R_{sd}^{D^*}$ defined in \cref{eq:ratdef}.
For $\ell = \mu$, we find
\begin{equation} \label{eq:ratios_vals}
    R_{sd}^D = 1.039 \pm  0.044 \ , \qquad
    R_{sd}^{D^*} = 0.991 \pm 0.036 \ ,
\end{equation}
that present relative uncertainties of $4.2\%$ and $3.6\%$ for the pseudoscalar and vector modes, respectively.
These estimates are compatible within uncertainties with the general expectation of $R_{sd}^{D^{(*)}} = 1 + \mathcal{O}(\%)$ discussed above.
For completeness, we give the correlation matrix
\begin{equation}
\rm{Corr} = \begin{pmatrix}
    1.0 & 0.0225 \\ 
   0.0225 &  1.0 \ 
\end{pmatrix} \ ,
\label{eq:ratios_correlation}
\end{equation}
which shows that the two determinations are currently very weakly correlated.
However, we note that correlations between the form factors for different modes are largely unknown, with the exception of the HPQCD $\bar B_{(s)}\to D_{(s)}^*\ell\bar\nu$ analysis and the LCSR results, which provide correlations between $\bar B\to D$ and $\bar B\to D^*$ form factors as well as those for the corresponding $\bar B_s$ transitions.
A dedicated lattice determination of the form factors correlated across different modes would therefore be very useful.

\begin{figure}
\centering
\begin{minipage}{.5\textwidth}
  \centering
  \includegraphics[width=.95\linewidth]{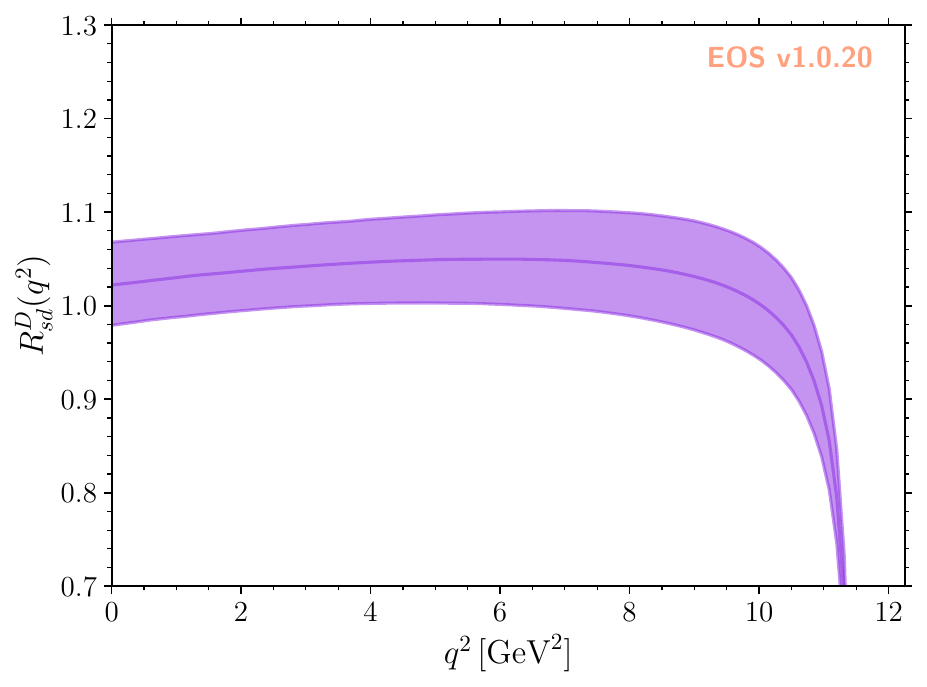}
\end{minipage}%
\begin{minipage}{.5\textwidth}
  \centering
  \includegraphics[width=.95\linewidth]{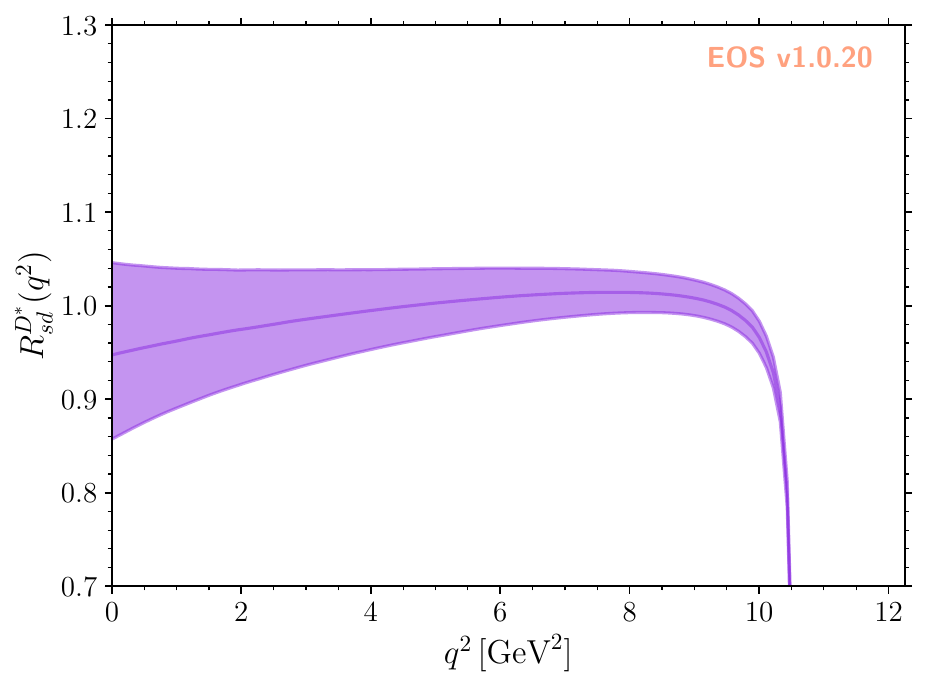}
\end{minipage}
\caption{Ratios $R_{sd}^D$ and $R_{sd}^{D^*}$ as functions of the squared momentum transfer $q^2$. The filled area corresponds to the $68\%$ probability envelope. The dotted lines indicate the average and $1\sigma$ interval for the $q^2$-integrated predictions of \cref{eq:ratios_vals}.}
\label{fig:diffRsd}
\end{figure}

In \cref{fig:diffRsd}, we display predictions for the ratios $R_{sd}^{D}$ and $R_{sd}^{D^*}$ as functions of the momentum transfer to the lepton-neutrino pair.
For the pseudoscalar mode, the relative uncertainty is almost constant and does not exceed $6\%$. 
For the vector mode, however, we find a clear $q^2$ dependence of the uncertainty.
It reaches $10\%$ at low $q^2$, where the form factors are mostly constrained by the LCSR estimates, while the uncertainty is only around $2\%$ at larger $q^2$ values, where most lattice data are determined and agree with each other~\cite{Bordone:2025jur}.
It might therefore be advantageous to perform a measurement in this area where theoretical uncertainties are reduced.
Estimates of the ratio in different $q^2$ bins can be readily obtained from our analysis.

\section{Towards $f_s/f_d$ from exclusive $\bar B_{(s)}\to D_{(s)}^{(*)}$ semi-leptonic decays } \label{sec:method}
We propose to extract $f_s/f_d$ from the measurement of the ratios of exclusive decays $R_{sd}^{D^{(*)}}$, in combination with the theoretical predictions in \cref{eq:ratios_vals}. 
The theoretical uncertainty of such an extraction is currently $4.2\% \;(3.6\%)$ for the $D^{(*)}$ modes.
Compared to previous determinations, this extraction has the advantage that theoretical uncertainties are controlled by lattice determinations, which are expected to improve further in the near future.

Below we analyse the current status of such an evaluation with a measurement aimed at determining $|V_{cb}|$ from $B_s$ decays \cite{LHCb:2020cyw}, see also Ref.~\cite{LHCb:2020hpv}.
Obtaining a precise value for $f_s/f_d$ utilising the upcoming lattice determinations of the corresponding form factors requires a measurement of the $R_{sd}^{D^{(*)}}$ ratios in a dedicated analysis.
As discussed, such an analysis could also investigate precision across different $q^2$ bins.
Finally, we note that while we focus here on semi-leptonic decays with muons, theoretical predictions for electron and tau final states are easily obtained as well.

\subsection{Proof of principle from current data}
To get an idea of the experimental precision for the ratios $R_{sd}^{D^{(*)}}$ and the corresponding constraint on $f_s/f_d$, we consider Ref.~\cite{LHCb:2020cyw} as a means to extract $f_s/f_d$.
In this analysis, the ratio of $\bar B_s\to D_s^{(*)}\ell\nu$ versus $\bar B_d\to D^{(*)}\ell\nu$ was studied using data taken at 7 and 8 TeV in order to extract $V_{cb}$ \cite{LHCb:2020cyw}. 
In the analysis, the ratio of branching ratios $\hat{R}_{sd}^{D^{(*)}}$, defined similarly to the ratios of decay rates in \cref{eq:ratdef}, are determined from a fit to
\begin{equation}
    N_{\rm sig}^{(*)} = N_{\rm ref}^{(*)}  \;\xi^{(*)}  \;\mathcal{K}^{(*)}  \;\hat{R}_{sd}^{D^{(*)}}  \ ,
\end{equation}
where $N_{\rm sig}$ are the signal $B_s^0$ decays, $N_{\rm ref}$ the reference $B_d^0$ decays, $\xi$ the efficiency ratio between signal and reference decays and 
\begin{equation}
    \mathcal{K}^{(*)} \equiv \frac{f_s}{f_d} \mathcal{R}_{D}^{(*)} \ ,
\end{equation}
where, for later use, we have defined
\begin{equation}
  \mathcal{R}_D \equiv \frac{\mathcal{B}(D_s^-\to K^+K^-\pi^-)}{\mathcal{B}(D^-\to K^+K^-\pi^-)}
  \ , \quad\quad\quad \mathcal{R}_D^* \equiv \frac{\mathcal{B}(D_s^-\to K^+K^-\pi^-)}{\mathcal{B}(D^-\to K^+K^-\pi^-) \, \mathcal{B}(D^{*-}\to D^- X)}  \ .
\end{equation}
The ratio $\mathcal{K}^{(*)}$ is an external input taken from Ref.~\cite{LHCb:2021qbv}. 

To obtain an estimate of $f_s/f_d$, we reconstruct the values of $N_{\rm sig}^{(*)} / N_{\rm ref}^{(*)} \xi^{(*)}$ based on the fit results and correlations of Tables 12 and 13 of Ref.~\cite{LHCb:2020cyw}.
We construct a multivariate normal distribution from the covariance matrix for
\begin{equation*}
\left\{ \frac{f_s}{f_d} \mathcal{B}(D_s^-\to K^+K^-\pi^-), \quad \mathcal{B}(D^-\to K^+K^-\pi^-), \quad \mathcal{B}(D^{*-}\to D^- X), \quad \hat{R}_{sd}^{D} , \quad \hat{R}_{sd}^{D^*} \right\},
\end{equation*}
and generate $10^6$ samples from which we obtain $N_{\rm sig} / N_{\rm ref} \xi = 1.40 \pm 0.10$  and $N_{\rm sig}^{*} / N_{\rm ref}^{*} \xi^{*} = 4.19\pm 0.34$, with a correlation of $-13\%$.

We then estimate $f_s/f_d$ using
\begin{equation}
    \left.\frac{f_s}{f_d}\right|_{D^{(*)}} = \frac{N_{\rm sig}^{(*)}}{N_{\rm ref}^{(*)} \xi^{(*)}} \frac{1}{\mathcal{R}_D^{(*)}}\frac{1}{R_{sd}^{D^{(*)}}} \frac{\tau_{B_d}}{\tau_{B_s}} \ .
\end{equation}
For the external inputs, we use the PDG values $\mathcal{B}(D_s^-\to K^+K^-\pi^-) = (5.37 \pm 0.10)\%$ and $\mathcal{B}(D^-\to K^+K^-\pi^-) = (9.68 \pm 0.18)\times 10^{-3}$\cite{ParticleDataGroup:2024cfk}.
For $\mathcal{B}(D^{*-}\to D^- X)$, we use the sum of the $D\pi^0$ and $D \gamma$ modes as in Ref.~\cite{LHCb:2020cyw}.
For the lifetime ratio, we use $\frac{\tau_{B_s}}{\tau_{B_d}} = 1.0021(34)$ \cite{HeavyFlavorAveragingGroupHFLAV:2024ctg}. 

Using the theory values for $R_{sd}^{D^{(*)}}$ given in \cref{eq:ratios_vals,eq:ratios_correlation} combined with our extracted ratios $\frac{N_{\rm sig}^{(*)}}{N_{\rm ref}^{(*)} \xi^{(*)}}$,
we again construct $10^6$ samples based on a multivariate normal distribution following the covariance matrix for all of our inputs.
Here, we assume all inputs to be uncorrelated and only consider the correlations between the $D$ and $D^*$ modes for the extracted number of events and the theory prediction.
This way we obtain
\begin{equation}
    \left.\frac{f_s}{f_d}\right|_D = 0.242\pm 0.022 \quad\mbox{and}\quad \left.\frac{f_s}{f_d}\right|_{D^*} =0.246 \pm 0.023\,,
\end{equation}
with a very small correlation, averaging to 
\begin{equation}\label{eq::resfsofd}
\left.\frac{f_s}{f_d}\right|_{D,D^*} = 0.244 \pm 0.016\,.
\end{equation}

This estimate has a $6.6\%$ precision and is in agreement with the global average of previous determinations, see \cref{eq:fsfd}.
While its uncertainty is about twice that of the previous global average, it is already competitive with the single determinations entering that average (see, e.g., the 7 TeV determination in Ref.~\cite{LHCb:2011leg}).
We emphasise that this result serves only as a proof of concept, highlighting the interesting prospects of this determination already with current data sets
and theoretical inputs.

\section{NP pollution} \label{sec:NPpollution}
Finally, we also consider the possible effects of heavy NP contributions in the charged-current decays under consideration. The aim of this section is not to perform a complete NP analysis of the charged-current operators, as done, e.g., in Refs.~\cite{Meiser:2024zea,Iguro:2020cpg,Jung:2018lfu,Carvunis:2025vab,Fael:2022wfc},
but rather to show that the extraction of the SM parameter $f_s/f_d$ is not polluted significantly by these potential NP contributions.

In the \SU symmetry limit, also the NP contributions cancel in the ratios $R_{sd}^{D^{(*)}}$, since the corresponding operators transform as singlets under the symmetry.
To further quantify this cancellation, we consider the effective $b\to c\ell\bar{\nu}_\ell$ Hamiltonian
\begin{equation}
    \mathcal{H}_{\text{eff}} = \frac{4 G_F V_{cb}}{\sqrt{2}} 
    \sum_i C_{i} \, \mathcal{O}_{i}\,,
    \qquad i \in \{V_L, V_R, S_L, S_R, T\},
    \label{eq::Hamiltonian-NP}
\end{equation}
with
\begin{align}
    \mathcal{O}_{V_{L(R)}} &= \left( \bar{c} \gamma_\mu P_{L(R)} b \right) \left(\bar{\ell} \gamma^\mu P_{L} \nu_\ell \right) , \nonumber \\
    \mathcal{O}_{S_{L(R)}} &= \left(\bar{c} P_{L(R)} b \right) \left(\bar{\ell} P_{L} \nu_\ell \right) , \\
    \mathcal{O}_{T} &= \left(\bar{c} \, \sigma_{\mu \nu} P_{L} b \right) \left(\bar{\ell} \, \sigma^{\mu \nu} P_{L} \nu_\ell \right) , \nonumber
\end{align}
and where $P_{L(R)} = \frac{1}{2}(1 \mp \gamma_5)$ and $\sigma^{\mu \nu} = \frac{i}{2} [\gamma^\mu, \gamma^\nu]$.
In the SM, $C_{V_L}^\mathrm{SM} = 1 + \frac{\alpha_e}{\pi} \ln \frac{M_Z}{\mu}$, and all other Wilson coefficients are zero.
Note that for this estimate of NP effects, we do not consider interactions with right-handed neutrinos (see, e.g., Ref.~\cite{Mandal:2020htr}).
The tree-level matching of SMEFT operators to the effective Hamiltonian can be found in Ref.~\cite{Aebischer:2015fzz}.

Using the above setup, we can quantify the NP pollution on the semi-leptonic ratios as
\begin{align} \label{eq:RsdNP}
    \frac{R_{sd}^{D}}{R_{sd,\mathrm{SM}}^{D}} & =
    \frac{\left|C_{V_L} + C_{V_R}\right|^2 + \eta_S^{D_s} \left|C_S\right|^2 + \eta_T^{D_s} \left|C_T\right|^2}{\left|C_{V_L} + C_{V_R}\right|^2 + \eta_S^{D} \left|C_S\right|^2 + \eta_T^{D} \left|C_T\right|^2} \\
    & \approx 1 + (\eta_S^{D_s} - \eta_S^{D}) \left|\tilde C_S\right|^2 + (\eta_T^{D_s} - \eta_T^{D}) \left|\tilde{C_T}\right|^2\,,
\end{align}
\begin{align} \label{eq:RsdstNP}
\frac{R_{sd}^{D^*}}{R_{sd,\mathrm{SM}}^{D^*}} & =
\frac{\left|C_{V_L}\right|^2 + \left|C_{V_R}\right|^2 + \eta_{V_LV_R}^{D_s^*} \mathrm{Re}\left(C_{V_L}^* C_{V_R}\right) + \eta_P^{D_s^*} \left|C_P\right|^2 + \eta_T^{D_s^*} \left|C_T\right|^2}
{\left|C_{V_L}\right|^2 + \left|C_{V_R}\right|^2 + \eta_{V_LV_R}^{D^*} \mathrm{Re}\left(C_{V_L}^* C_{V_R}\right) + \eta_P^{D^*} \left|C_P\right|^2 + \eta_T^{D^*} \left|C_T\right|^2} \\
& \approx 1 + (\eta_P^{D_s^*} - \eta_P^{D^*})\left|\tilde C_P\right|^2 + (\eta_T^{D_s^*} - \eta_T^{D^*})\left|\tilde C_T\right|^2 \nonumber \\
& \quad +(\eta_{V_LV_R}^{D_s^*} - \eta_{V_LV_R}^{D^*})\,\mathrm{Re}\left(\tilde C_{V_R}\right)\left[1 - \eta_{V_LV_R}^{D^*} \mathrm{Re}\left(\tilde C_{V_R}\right)\right]\,,
\end{align}
where $\tilde C_i = C_i / C_{V_L}$, $C_{S(P)} = C_{S_L} \pm C_{S_R}$, and we neglected the lepton mass for simplicity.
The coefficients $\eta_i$ depend on ratios of the hadronic form factors and the masses in the transition \cite{Becirevic:2012jf,Duraisamy:2013pia}.
Their values (for $m_\mu=0$) are summarised in \cref{tab:NPpollution} and available in full generality as supplementary material, as described in \cref{sec:supplementary}.

Due to the limited theory inputs on the tensor form factor $f_T$ for the $B\to D$ and $B_s\to D_s$ transitions, the coefficients $\eta_T^{D}$ and $\eta_T^{D_s}$ are currently poorly constrained.
Their size can be estimated using the HQET relations between $f_T$ and other $\bar B_{(s)}\to D_{(s)}$ form factors.
For the present pilot BSM analysis, we simply extract synthetic $f_T$ data points from the global HQET analysis of Ref.~\cite{Bordone:2025jur} and use them as additional uncorrelated constraints.

\begin{table}[t]
    \centering
    \begin{tabular}{ccc}
        \toprule
        $\eta_S^D$      & $\eta_S^{D_s}$    & $\eta_S^{D_s} - \eta_S^D$ \\
        $1.13 \pm 0.01$ & $1.14 \pm 0.03$   & $(0.8 \pm 3.5) \, 10^{-2}$ \\
        \midrule
        $\eta_T^D$      & $\eta_T^{D_s}$    & $\eta_T^{D_s} - \eta_T^D$ \\
        $0.76 \pm 0.04$ & $0.78 \pm 0.08$   & $(2.9 \pm 9.2) \, 10^{-2}$ \\
        \midrule
        $\eta_{V_LV_R}^{D^*}$   & $\eta_{V_LV_R}^{D_s^*}$   & $\eta_{V_LV_R}^{D_s^*} - \eta_{V_LV_R}^{D^*}$ \\
        $-1.75 \pm 0.01$        & $-1.75 \pm 0.02$          & $(0.5 \pm 1.9) \, 10^{-2}$ \\
        \midrule
        $\eta_P^{D^*}$      & $\eta_P^{D_s^*}$  & $\eta_P^{D_s^*} - \eta_P^{D^*}$ \\
        $0.067 \pm 0.002$   & $0.067 \pm 0.003$ & $(0.1 \pm 2.7) \, 10^{-3}$ \\
        \midrule
        $\eta_T^{D^*}$  & $\eta_T^{D_s^*}$  & $\eta_T^{D_s^*} - \eta_T^{D^*}$ \\
        $17.2 \pm 0.9$  & $17.5 \pm 0.8$    & $0.22 \pm 0.90$ \\
        \bottomrule
    \end{tabular}
    \caption{\label{tab:NPpollution}
        Values of the relative operator corrections to the SM rates, in the limit of massless leptons, as defined in \cref{eq:RsdNP,eq:RsdstNP}.
        The general case and the correlations are provided in the supplementary materials, as described in \cref{sec:supplementary}.
    }
\end{table}

The \SU-breaking effects of the $\eta_i$ parameters are all consistent with zero, in accordance with the general expectation.
We find that relative \SU-breaking of the order of a few per cent remains allowed by the current lattice data, with the largest potential effects in the tensor contributions.
Independent constraints on the norm of the reduced Wilson coefficients $\tilde C_i \equiv C_i / C_{V_L}$ can be obtained from global analyses of inclusive observables~\cite{Carvunis:2025vab}.
In that reference, the coefficients are found to be $|\tilde C_{V_R}| = 0.19^{+0.21}_{-0.16}$, $|\tilde C_S| = 0.45^{+0.10}_{-0.14}$, $|\tilde C_P| =0.20^{+0.43}_{-0.20}$ and $|\tilde C_T| = 0.15^{+0.07}_{-0.11}$.
However, the corresponding posterior distributions are not Gaussian.
In order to estimate the allowed contributions to $R_{sd}^{D^{(*)}}$, we use directly the marginalised likelihoods plotted in the paper.
We find that all contributions to $R_{sd}^{D^{(*)}}$ are limited to be at most at the per-mil level, except for the tensor ones in $\bar B\to D_{(s)}^*$, for which we obtain
\begin{equation}
    (\eta_T^{D_s^*} - \eta_T^{D^*})\left(\frac{C_T}{C_{V_L}}\right)^2 \in [-1.2\%,2.1\%],
\end{equation}
at $68\%$ confidence level.

We conclude that all heavy NP contributions are below the current experimental sensitivity, with most of them constrained to be smaller by about a factor of ten or more.
Future improvements in the calculation of the $\bar B\to D_{(s)}^*$ tensor form factors or in the measurement of inclusive moments will further reduce uncertainties in these contributions.

\section{Generalisation of the method} \label{sec:fcfd}
The extraction of fragmentation functions from exclusive semi-leptonic modes can be generalised to other modes, provided that both experimental data and theoretical predictions of the form factors are available.
One promising example is the determination of the relative \Bc production fraction $f_c/f_{d,u}$.
The LHCb collaboration extracted this ratio from the ratio of $B_c^+\to J/\psi\mu^+\nu_\mu$ with respect to the semi-inclusive $B\to D^{\pm,0}X\mu\nu_\mu$ rates, see also the discussion in Ref.~\cite{Amhis:2021cfy}.
Using the experimental average of the inclusive rate for $B\to X\mu\nu$ from Ref.~\cite{LHCb:2016qpe} which combines the results from \cite{Belle:2006kgy, BaBar:2006ztv, CLEO:2004stg} together with \cref{eq::BDXapprox},  
the result reads \cite{LHCb:2019tea}
\begin{equation}\label{eq:fcfufd}
  \frac{f_c}{f_u + f_d}  \mathcal{B}(B_c^+\to J/\psi\mu^+\nu_\mu) = (7.36 \pm 0.08\pm 0.30)\cdot 10^{-5} \,,
\end{equation}
where the production fractions are those at 13~TeV.
We note that this normalisation with respect to the semi-inclusive mode will again be affected by the issues detailed above.
The $B_c\to J/\psi$ form factors have been calculated on the lattice by the HPQCD collaboration \cite{Harrison:2020gvo,Harrison:2025yan}.
Using $|V_{cb}| =  (41.82^{+0.85}_{-0.74})\cdot 10^{-3}$, they obtain
\begin{equation}
    \mathcal{B}(B_c^+\to J/\psi\mu^+\nu_\mu) = (1.53 \pm 0.08)\%  \ .
\end{equation}
Combining this with \cref{eq:fcfufd} yields
\begin{equation}
     \frac{f_c}{f_u + f_d} = (4.81 \pm 0.32) \cdot 10^{-3}.
\end{equation}

We propose to determine $f_c/f_d$ instead from measurements of $B_c^+\to J/\psi \mu^+ \nu_\mu$ relative to $B\to D^*\mu^+\nu_\mu$ decays.
Using theory expressions for both numerator and denominator has again the advantage that some theory uncertainties cancel, most importantly the dependence on $|V_{cb}|$.
Furthermore, choosing both modes as pseudoscalar-to-vector transitions maximises the cancellation of possible NP effects.
As discussed above, lattice QCD predictions for the form factors of both modes are available, albeit without correlations between the different modes.
A first exploration using QCD sum rules is also available \cite{Bordone:2022drp}, and we look forward to new determinations of these form factors.

\begin{figure}
    \centering
    \includegraphics[width=0.48\linewidth]{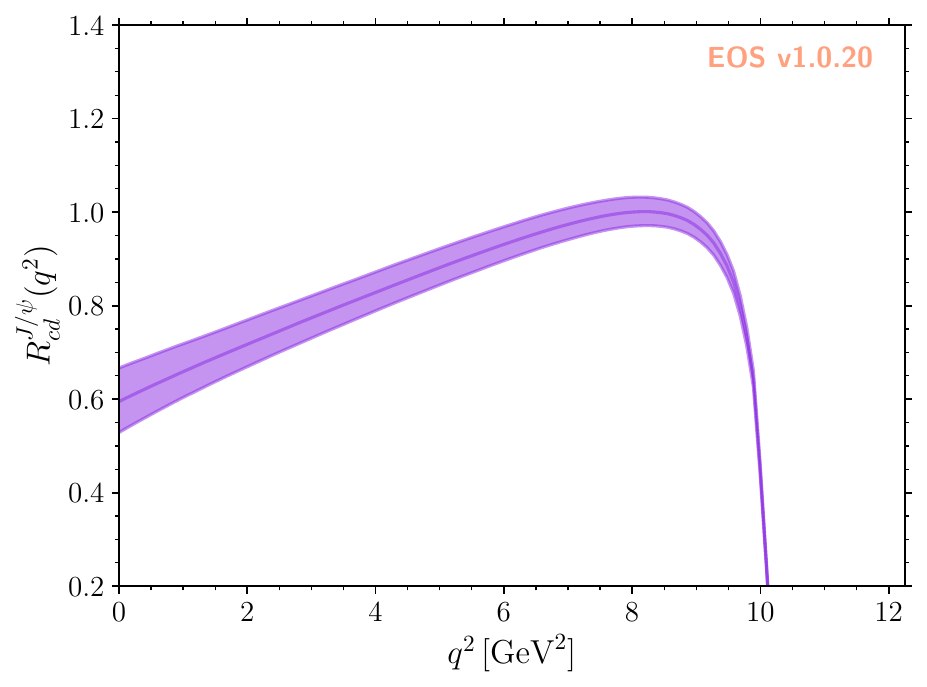}
    \caption{Ratio $R_{cd}^{J/\psi}$ as function of the squared momentum transfer $q^2$. The filled area corresponds to the $68\%$ probability envelope. The dotted lines indicate the average and $1\sigma$ interval for the $q^2$-integrated prediction of \cref{eq:ratiojpsi}.}
    \label{fig:diffRcd}
\end{figure}

Using only the LQCD results of Ref.~\cite{Harrison:2025yan} and neglecting their correlations with the other modes, we obtain
\begin{equation} \label{eq:ratiojpsi}
    R_{cd}^{J/\psi} \equiv \frac{\Gamma(B_c^+ \to J/\psi\mu^+\nu_\mu)}{\Gamma(B_d\to D_d^{*-} \mu^+ \nu_\mu)} = 0.832 \pm 0.036 \,,
\end{equation}
which has a $4\%$ relative uncertainty.
We also look forward to an experimental determination of this ratio.
The corresponding differential ratio is shown in \cref{fig:diffRcd}.
We find this ratio to be more sensitive to phase space effects than $R_{sd}^{D^{(*)}}$.
For completeness, we also provide
\begin{equation}
    \frac{\Gamma(B_c^+\to J/\psi\mu^+\nu_\mu)}{\Gamma(B_d\to D_d^- \mu^+\nu_\mu)} =  1.895 \pm 0.070 \,,
\end{equation}
but we expect this ratio to be more sensitive to BSM pollution.\\

Another possible application of our method is the determination of the relative baryon production fraction $f_{\Lambda_b} / f_d$.
The current determination is again based on semi-inclusive decays \cite{LHCb:2011leg, LHCb:2019fns} and hence suffers from the same issues as discussed above for $f_s$.
At $pp$ colliders, the branching ratio of the semi-leptonic decay $\Lambda_b\to\Lambda_c\ell\nu_\ell$ has only been measured for $\ell = \tau$~\cite{LHCb:2022piu} or with respect to $\Lambda_b\to p \mu\nu_\mu$~\cite{LHCb:2015eia}, preventing us from providing a first determination with our method as for $f_s/f_d$.
The main difficulty in applying our method to $\Lambda_b$ lies in obtaining precise predictions for the form factors. On the lattice, this is significantly more difficult due to a higher signal-to-noise ratio.
First results for the total rate with relativistic heavy-quark actions read $\Gamma(\Lambda_b\to\Lambda_c^+\mu^-\nu_\mu) / |V_{cb}|^2 = (21.5 \pm 0.8 \pm 1.1)\,\mathrm{ps}^{-1}$ \cite{Detmold:2015aaa}, which hopefully will be improved upon in the future.
We also note that by normalising baryonic to mesonic modes, we do not expect a good cancellation of possible NP effects.
Nevertheless, a partial cancellation can be achieved, for instance for the tensor contribution, while improved limits from other modes, such as the inclusive one used above, will help to keep the NP pollution under control.

\section{Conclusion}

We propose extracting ratios of production fractions in a systematically improvable approach by predicting and measuring ratios of exclusive semi-leptonic decays.
The advantages of this method over the current state-of-the-art extractions are the following:
\begin{enumerate}
    \item[(i)] The theory uncertainties are dominated by hadronic form factors, and can therefore be systematically reduced in the future.
    In the case of $f_s/f_d$, the corresponding form factors only differ by additionally suppressed \SU-breaking effects, and are hence expected to be particularly small.
    \item[(ii)] Thanks to the similarities between the decays in the numerator and denominator, experimental systematic uncertainties partially cancel as well.
    \item[(iii)] The method is robust against potential pollution from heavy NP contributions, as they partially cancel in the ratio.
\end{enumerate} 

For the determination of $f_s/f_d$, we suggest to use the ratios of $\Bsb\to D_s^{(*)}\mu\bar\nu$ and $\Bdb\to D^{(*)}\mu\bar\nu$ decays.
To demonstrate this new method, we extract $f_s/f_d$ from a previous LHCb analysis using 7 and 8 TeV data~\cite{LHCb:2020cyw}, employing the current state-of-the-art lattice and sum rule determinations of the required form factors.
We predict $R_{sd}^{D^{(*)}}$ with $4.6\% \; (3.8\%)$ relative uncertainty, and obtain a $7\%$ uncertainty for $f_s/f_d$, see \cref{eq::resfsofd}.
We also explicitly demonstrate that the maximum NP pollution in this analysis is significantly below the current uncertainty.

It is important to note that the theoretical uncertainties in our approach could already be further reduced with the available data: in this work, we constrain the form factors only from theory inputs, and the NP pollution only from inclusive decays.
Both sets of constraints could be improved by performing a global fit to all available semi-leptonic data, including, in particular, the experimental information on differential rates.
This more involved approach would also leverage the different sensitivity to $f_s/f_d$ in different $q^2$ bins discussed above (all our formulas can be trivially extended to the full set of $q^2$-dependent angular observables).
Such an analysis is beyond the scope of the present work, in particular, because the theory uncertainties are subleading in our present extraction of $f_s/f_d$.
Once improved measurements of $R_{sd}^D$ and $R_{sd}^{D^*}$ become available, this analysis can and should be revisited, incorporating potential further improvements, such as correlated lattice determinations.

Finally, we note that our method can also be applied to other $b$ hadrons, like $\Lambda_b$ or \Bc.
Interesting prospects arise in particular for the extraction of the ratio $f_c/f_d$ from $R_{cd}^{J/\psi}$, with a current theoretical uncertainty of $4\%$.

\acknowledgments
We thank M. Rotondo and J. Ruiz Vidal for useful discussions. This publication is part of the project Solving Beautiful Puzzles with file number VI.Vidi.223.083 of the research programme Vidi which is financed by the Dutch Research Council (NWO). C.B. appreciates support by the Deutsche Forschungsgemeinschaft (DFG, German Research Foundation) under Germany’s Excellence Strategy – EXC 3107 – Project-ID 533766364 and within the Emmy Noether Program under grant number MI 2869/1-1.
M.J. acknowledges support by the Italian Ministry of University and Research (MUR) under grant PRIN 2022N4W8WR.
\appendix

\section{Form factor analysis} \label{app:form-factor-constraints}
In our form factor analysis, we considered the following constraints:
\begin{itemize}
    \item $B\to D$ LQCD: The $f_0$ and $f_+$ form factors have been estimated by the HPQCD \cite{Na:2015kha} and the FNAL+MILC \cite{FermilabLattice:2015ilb} collaborations.
    \item $B_s\to D_s$ LQCD: The $f_0$ and $f_+$ form factors have been estimated by the HPQCD collaboration \cite{McLean:2019qcx}.
    \item $B\to D^*$ LQCD: The vector and axial form factors have been estimated by the FNAL+MILC \cite{FermilabLattice:2021cdg} and the JLQCD collaborations \cite{Aoki:2023qpa}.
    \item $B_{(s)} \to D_{(s)}^*$ LQCD: The complete set of form factors has been estimated by the HPQCD collaboration \cite{Harrison:2023dzh}.
    \item $B\to D^{(*)}$ LCSR: The complete set of form factors has been estimated in Ref. \cite{Gubernari:2018wyi}.
    \item $B_s\to D_s^{(*)}$ LCSR: The complete set of form factors has been estimated in Ref. \cite{Bordone:2019guc}
    \item $B_c\to J/\psi$ LQCD: The complete set of form factors has been estimated by the HPQCD collaboration \cite{Harrison:2025yan}.
\end{itemize}
Our fit results and the plot of all the form factors are available in the supplementary materials~\cite{EOS-DATA-2026-02}.
We also compute the saturations of the dispersive bounds \textit{a posteriori}.
At the truncation order at which we are working, $N = 2$, we find the following $95\%$ probability intervals for the saturations
\begin{align}
    \chi_{0^+_V} < 0.084 & \qquad \chi_{0^-_A} < 2.5\\
    \chi_{1^-_V} < 0.50  & \qquad \chi_{1^+_A} < 0.035 \\
    \chi_{1^-_T} < 0.077 & \qquad \chi_{1^+_{T5}} < 0.069
\end{align}
We conclude that all the bounds are satisfied, except for the pseudoscalar bound.
For the latter, we find that the dominant contribution comes from $B\to D^*$ transitions.
We find that restricting our samples to those that satisfy all the bounds has a negligible impact on our analysis results.

\section{Reproducing our results} \label{sec:supplementary}
To facilitate the use of our results, we provide the code used for this analysis in the analysis repository~\cite{EOS-DATA-2026-02}.
Running the analysis requires installing \EOS version 1.0.20 or newer.

The repository also contains additional figures for the fit to the form factors, as well as the necessary ingredients to study the impact of NP in the non-vanishing lepton mass case.
For the latter, we write the four branching ratios $\mathcal{B}(B_{(s)} \to D_{(s)}^{(*)})$ in the form
\begin{equation}
    \mathcal{B} = \bar{\mathcal{C}} \mathcal{M} \mathcal{C} \qquad
    \mathcal{C} = \left\{ C_{V_L}, C_{V_R}, C_{S_L}, C_{S_R}, C_T \right\} \,,
\end{equation}
and we provide the means and covariances for $\left\{\frac{\mathcal{M}_{ij}}{\mathcal{M}_{00}}\right\}$ in the order $B_s\to D_s$, $B\to D$, $B_s\to D_s^*$ and $B\to D^*$.

\bibliographystyle{jhep}
\bibliography{references.bib}

@article{Isgur:1989vq,
    author = "Isgur, Nathan and Wise, Mark B.",
    title = "{Weak Decays of Heavy Mesons in the Static Quark Approximation}",
    reportNumber = "UTPT-89-27, CALT-68-1585",
    doi = "10.1016/0370-2693(89)90566-2",
    journal = "Phys. Lett. B",
    volume = "232",
    pages = "113--117",
    year = "1989"
}

@article{Bernlochner:2022ywh,
    author = "Bernlochner, Florian U. and Ligeti, Zoltan and Papucci, Michele and Prim, Markus T. and Robinson, Dean J. and Xiong, Chenglu",
    title = "{Constrained second-order power corrections in HQET: $R(D^{(*)})$, $|V_{cb}|$, and new physics}",
    eprint = "2206.11281",
    archivePrefix = "arXiv",
    primaryClass = "hep-ph",
    reportNumber = "CALT-TH-2022-022",
    doi = "10.1103/PhysRevD.106.096015",
    journal = "Phys. Rev. D",
    volume = "106",
    number = "9",
    pages = "096015",
    year = "2022"
}

@article{Martinelli:2023fwm,
    author = "Martinelli, G. and Simula, S. and Vittorio, L.",
    title = "{Updates on the determination of $\vert V_{cb} \vert ,R(D^{*})$ and $\vert V_{ub} \vert /\vert V_{cb} \vert $}",
    eprint = "2310.03680",
    archivePrefix = "arXiv",
    primaryClass = "hep-ph",
    doi = "10.1140/epjc/s10052-024-12742-5",
    journal = "Eur. Phys. J. C",
    volume = "84",
    number = "4",
    pages = "400",
    year = "2024"
}

@article{Bordone:2024weh,
    author = "Bordone, Marzia and Juttner, Andreas",
    title = "{New strategies for probing $B\rightarrow D^*\ell \bar{\nu }_\ell $ lattice and experimental data}",
    eprint = "2406.10074",
    archivePrefix = "arXiv",
    primaryClass = "hep-ph",
    reportNumber = "CERN-TH-2024-083",
    doi = "10.1140/epjc/s10052-025-13773-2",
    journal = "Eur. Phys. J. C",
    volume = "85",
    number = "2",
    pages = "129",
    year = "2025"
}

@article{Fedele:2023ewe,
    author = "Fedele, Marco and Blanke, Monika and Crivellin, Andreas and Iguro, Syuhei and Nierste, Ulrich and Simula, Silvano and Vittorio, Ludovico",
    title = "{Discriminating B{\textrightarrow}D*{\ensuremath{\ell}}{\ensuremath{\nu}} form factors via polarization observables and asymmetries}",
    eprint = "2305.15457",
    archivePrefix = "arXiv",
    primaryClass = "hep-ph",
    reportNumber = "PSI-PR-23-15, PSI-PR-23-12, ZU-TH 22/23, TTP23-019, P3H-23-033, LAPTH-020/23",
    doi = "10.1103/PhysRevD.108.055037",
    journal = "Phys. Rev. D",
    volume = "108",
    number = "5",
    pages = "055037",
    year = "2023"
}

@article{Ray:2023xjn,
    author = "Ray, Ipsita and Nandi, Soumitra",
    title = "{Test of new physics effects in $ \overline{B}\to \left({D}^{\left(\ast \right)},\pi \right){\ell}^{-}{\overline{\nu}}_{\ell } $ decays with heavy and light leptons}",
    eprint = "2305.11855",
    archivePrefix = "arXiv",
    primaryClass = "hep-ph",
    doi = "10.1007/JHEP01(2024)022",
    journal = "JHEP",
    volume = "01",
    pages = "022",
    year = "2024"
}

@article{Gopal:2024mgb,
    author = "Gopal, Abinand and Gubernari, Nico",
    title = "{Unitarity bounds with subthreshold and anomalous cuts for b-hadron decays}",
    eprint = "2412.04388",
    archivePrefix = "arXiv",
    primaryClass = "hep-ph",
    doi = "10.1103/PhysRevD.111.L031501",
    journal = "Phys. Rev. D",
    volume = "111",
    number = "3",
    pages = "L031501",
    year = "2025"
}

@article{Belle:2002lms,
    author = "Hastings, N. C. and others",
    collaboration = "Belle",
    title = "{Studies of $B^0-\bar{B}^0$ mixing properties with inclusive dilepton events}",
    eprint = "hep-ex/0212033",
    archivePrefix = "arXiv",
    reportNumber = "KEK-PREPRINT-2002-123, BELLE-PREPRINT-2002-38",
    doi = "10.1103/PhysRevD.67.052004",
    journal = "Phys. Rev. D",
    volume = "67",
    pages = "052004",
    year = "2003"
}

@article{CLEO:1990fzo,
    author = "Fulton, R. and others",
    collaboration = "CLEO",
    title = "{Exclusive and inclusive semileptonic decays of B mesons to D mesons}",
    reportNumber = "CLNS-90-989, CLEO-90-4",
    doi = "10.1103/PhysRevD.43.651",
    journal = "Phys. Rev. D",
    volume = "43",
    pages = "651--663",
    year = "1991"
}

@article{BaBar:2007xlq,
    author = "Aubert, Bernard and others",
    collaboration = "BaBar",
    title = "{Measurement of the relative branching fractions of $\bar{B}\to D/D^*/D^{**}\ell^-\bar{\nu}_\ell$ decays in events with a fully reconstructed $B$ meson}",
    eprint = "hep-ex/0703027",
    archivePrefix = "arXiv",
    reportNumber = "SLAC-PUB-12393, BABAR-PUB-07-012",
    doi = "10.1103/PhysRevD.76.051101",
    journal = "Phys. Rev. D",
    volume = "76",
    pages = "051101",
    year = "2007"
}

@article{Boyd:1995sq,
    author = "Boyd, C. Glenn and Grinstein, Benjamin and Lebed, Richard F.",
    title = "{Model independent determinations of $\bar{B}\to D\ell, D^*\ell\bar{\nu}$ form-factors}",
    eprint = "hep-ph/9508211",
    archivePrefix = "arXiv",
    reportNumber = "UCSD-PTH-95-11",
    doi = "10.1016/0550-3213(95)00653-2",
    journal = "Nucl. Phys. B",
    volume = "461",
    pages = "493--511",
    year = "1996"
}

@article{Caprini:1995wq,
    author = "Caprini, Irinel and Neubert, Matthias",
    title = "{Improved bounds for the slope and curvature of $\bar{B} \to D^{(*)}$ lepton anti-neutrino form-factors}",
    eprint = "hep-ph/9603414",
    archivePrefix = "arXiv",
    reportNumber = "CERN-TH-95-255",
    doi = "10.1016/0370-2693(96)00509-6",
    journal = "Phys. Lett. B",
    volume = "380",
    pages = "376--384",
    year = "1996"
}

@article{Jung:2014jfa,
    author = "Jung, Martin and Schacht, Stefan",
    title = "{Standard model predictions and new physics sensitivity in $B \to DD$ decays}",
    eprint = "1410.8396",
    archivePrefix = "arXiv",
    primaryClass = "hep-ph",
    reportNumber = "DO-TH-13-24, TTP-14-027",
    doi = "10.1103/PhysRevD.91.034027",
    journal = "Phys. Rev. D",
    volume = "91",
    number = "3",
    pages = "034027",
    year = "2015"
}

@article{Bigi:2011gf,
    author = "Bigi, I. I. and Mannel, Th. and Uraltsev, N.",
    title = "{Semileptonic width ratios among beauty hadrons}",
    eprint = "1105.4574",
    archivePrefix = "arXiv",
    primaryClass = "hep-ph",
    reportNumber = "UND-HEP-11-BIG02",
    doi = "10.1007/JHEP09(2011)012",
    journal = "JHEP",
    volume = "09",
    pages = "012",
    year = "2011"
}

@article{Harrison:2024iad,
    author = "Harrison, Judd",
    title = "{$b\bar{c}$ susceptibilities from fully relativistic lattice QCD}",
    eprint = "2405.01390",
    archivePrefix = "arXiv",
    primaryClass = "hep-lat",
    doi = "10.1103/PhysRevD.110.054506",
    journal = "Phys. Rev. D",
    volume = "110",
    number = "5",
    pages = "054506",
    year = "2024"
}

@article{Bigi:2017jbd,
    author = "Bigi, Dante and Gambino, Paolo and Schacht, Stefan",
    title = "{$R(D^*)$, $|V_{cb}|$, and the Heavy Quark Symmetry relations between form factors}",
    eprint = "1707.09509",
    archivePrefix = "arXiv",
    primaryClass = "hep-ph",
    doi = "10.1007/JHEP11(2017)061",
    journal = "JHEP",
    volume = "11",
    pages = "061",
    year = "2017"
}

@article{Davies:2023arm,
    author = "Davies, Jonathan and Jung, Martin and Schacht, Stefan",
    title = "{$ \overline{B}\to \overline{D}D $ decays and the extraction of f$_{d}$/f$_{u}$ at hadron colliders}",
    eprint = "2311.16952",
    archivePrefix = "arXiv",
    primaryClass = "hep-ph",
    doi = "10.1007/JHEP01(2024)191",
    journal = "JHEP",
    volume = "01",
    pages = "191",
    year = "2024",
    note = "[Erratum: JHEP 01, 162 (2025)]"
}

@article{CMS:2026kkx,
    author = "Hayrapetyan, Aram and others",
    collaboration = "CMS",
    title = "{Measurement of B meson production fraction ratios in proton-proton collisions at $\sqrt{s}$ = 13 TeV using open-charm and charmonium decays}",
    eprint = "2602.10270",
    archivePrefix = "arXiv",
    primaryClass = "hep-ex",
    reportNumber = "CMS-BPH-21-007, CERN-EP-2025-176",
    month = "2",
    year = "2026"
}

@article{CMS:2022wkk,
    author = "Tumasyan, Armen and others",
    collaboration = "CMS",
    title = "{Measurement of the dependence of the hadron production fraction ratio $f_\mathrm{s} / f_\mathrm{u}$ and $f_\mathrm{d} / f_ \mathrm{u}$ on B meson kinematic variables in proton-proton collisions at $\sqrt{s}$ = 13 TeV}",
    eprint = "2212.02309",
    archivePrefix = "arXiv",
    primaryClass = "hep-ex",
    reportNumber = "CMS-BPH-21-001, CERN-EP-2022-248",
    doi = "10.1103/PhysRevLett.131.121901",
    journal = "Phys. Rev. Lett.",
    volume = "131",
    pages = "121901",
    year = "2023"
}

@article{LHCb:2013sad,
    author = "Aaij, R and others",
    collaboration = "LHCb",
    title = "{First observations of $\bar{B}_s^0 \to D^+D^-$, $D_s^+D^-$ and $D^0\bar{D}^0$ decays}",
    eprint = "1302.5854",
    archivePrefix = "arXiv",
    primaryClass = "hep-ex",
    reportNumber = "LHCB-PAPER-2012-050, CERN-PH-EP-2013-018",
    doi = "10.1103/PhysRevD.87.092007",
    journal = "Phys. Rev. D",
    volume = "87",
    number = "9",
    pages = "092007",
    year = "2013"
}

@article{LHCb:2018azb,
    author = "Aaij, Roel and others",
    collaboration = "LHCb",
    title = "{Measurement of the relative $B^{-} \!\rightarrow D^{0} / D^{*0} / D^{**0} \mu^{-} \overline{\nu}_\mu$ branching fractions using $B^{-}$ mesons from $\overline{B}{}_{s2}^{*0}$ decays}",
    eprint = "1807.10722",
    archivePrefix = "arXiv",
    primaryClass = "hep-ex",
    reportNumber = "LHCb-PAPER-2018-024, CERN-EP-2018-190",
    doi = "10.1103/PhysRevD.99.092009",
    journal = "Phys. Rev. D",
    volume = "99",
    number = "9",
    pages = "092009",
    year = "2019"
}

@article{Kobach:2019kfb,
    author = "Kobach, Andrew",
    title = "{Continuity and Semileptonic $B_{(s)}\rightarrow D_{(s)}$ Form Factors}",
    eprint = "1910.13024",
    archivePrefix = "arXiv",
    primaryClass = "hep-ph",
    doi = "10.1016/j.physletb.2020.135708",
    journal = "Phys. Lett. B",
    volume = "809",
    pages = "135708",
    year = "2020"
}

@article{Isgur:1990yhj,
    author = "Isgur, Nathan and Wise, Mark B.",
    title = "{Weak transition form factors between heavy mesons}",
    reportNumber = "UTPT-90-01, CALT-68-1608",
    doi = "10.1016/0370-2693(90)91219-2",
    journal = "Phys. Lett. B",
    volume = "237",
    pages = "527--530",
    year = "1990"
}

@article{Bordone:2022qez,
    author = "Bordone, Marzia and Gambino, Paolo",
    title = "{The semileptonic $B_s$ and $\Lambda_b$ widths}",
    eprint = "2203.13107",
    archivePrefix = "arXiv",
    primaryClass = "hep-ph",
    reportNumber = "CERN-TH-2022-050",
    doi = "10.22323/1.411.0055",
    journal = "PoS",
    volume = "CKM2021",
    pages = "055",
    year = "2023"
}

@article{Fleischer:2010ay,
    author = "Fleischer, Robert and Serra, Nicola and Tuning, Niels",
    title = "{A New Strategy for $B_s$ Branching Ratio Measurements and the Search for New Physics in $B^0_s \to \mu^+ \mu^-$}",
    eprint = "1004.3982",
    archivePrefix = "arXiv",
    primaryClass = "hep-ph",
    reportNumber = "NIKHEF-2010-010",
    doi = "10.1103/PhysRevD.82.034038",
    journal = "Phys. Rev. D",
    volume = "82",
    pages = "034038",
    year = "2010"
}

@article{Beneke:2000ry,
    author = "Beneke, M. and Buchalla, G. and Neubert, M. and Sachrajda, Christopher T.",
    title = "{QCD factorization for exclusive, nonleptonic B meson decays: General arguments and the case of heavy light final states}",
    eprint = "hep-ph/0006124",
    archivePrefix = "arXiv",
    reportNumber = "CERN-TH-2000-159, CLNS-00-1675, PITHA-00-06, SHEP-00-06",
    doi = "10.1016/S0550-3213(00)00559-9",
    journal = "Nucl. Phys. B",
    volume = "591",
    pages = "313--418",
    year = "2000"
}

@article{Huber:2016xod,
    author = {Huber, Tobias and Kr{\"a}nkl, Susanne and Li, Xin-Qiang},
    title = "{Two-body non-leptonic heavy-to-heavy decays at NNLO in QCD factorization}",
    eprint = "1606.02888",
    archivePrefix = "arXiv",
    primaryClass = "hep-ph",
    reportNumber = "SI-HEP-2016-12, QFET-2016-06",
    doi = "10.1007/JHEP09(2016)112",
    journal = "JHEP",
    volume = "09",
    pages = "112",
    year = "2016"
}

@article{Bordone:2020gao,
    author = "Bordone, Marzia and Gubernari, Nico and Huber, Tobias and Jung, Martin and van Dyk, Danny",
    title = "{A puzzle in $\bar{B}_{(s)}^0 \to D_{(s)}^{(*)+} \lbrace \pi^-, K^-\rbrace$ decays and extraction of the $f_s/f_d$ fragmentation fraction}",
    eprint = "2007.10338",
    archivePrefix = "arXiv",
    primaryClass = "hep-ph",
    reportNumber = "TUM-HEP 1271/20, P3H-20-034, SI-HEP-2020-17",
    doi = "10.1140/epjc/s10052-020-08512-8",
    journal = "Eur. Phys. J. C",
    volume = "80",
    number = "10",
    pages = "951",
    year = "2020"
}

@article{LHCb:2016qpe,
    author = "Aaij, Roel and others",
    collaboration = "LHCb",
    title = "{Measurement of the $b$-quark production cross-section in 7 and 13 TeV $pp$ collisions}",
    eprint = "1612.05140",
    archivePrefix = "arXiv",
    primaryClass = "hep-ex",
    reportNumber = "CERN-EP-2016-201, LHCB-PAPER-2016-031",
    doi = "10.1103/PhysRevLett.118.052002",
    journal = "Phys. Rev. Lett.",
    volume = "118",
    number = "5",
    pages = "052002",
    year = "2017",
    note = "[Erratum: Phys.Rev.Lett. 119, 169901 (2017)]"
}

@article{Bordone:2022drp,
    author = "Bordone, Marzia and Khodjamirian, Alexander and Mannel, Th.",
    title = "{New sum rules for the $B_c\to J/\psi$ form factors}",
    eprint = "2209.08851",
    archivePrefix = "arXiv",
    primaryClass = "hep-ph",
    reportNumber = "SI-HEP-2022-26, P3H-22-094, CERN-TH-2022-144",
    doi = "10.1007/JHEP01(2023)032",
    journal = "JHEP",
    volume = "01",
    pages = "032",
    year = "2023"
}

@article{Bordone:2025jur,
    author = "Bordone, Marzia and Gubernari, Nico and Jung, Martin and van Dyk, Danny",
    title = "{Challenging $ {\overline{B}}_{(s)}\to {D}_{(s)}^{\left(\ast \right)} $ form factors with the heavy quark expansion}",
    eprint = "2507.03569",
    archivePrefix = "arXiv",
    primaryClass = "hep-ph",
    reportNumber = "CERN-TH-2025-092, EOS-2025-03, IPPP/25/25, P3H-25-032, SI-HEP-2025-10, ZU-TH 35/25",
    doi = "10.1007/JHEP11(2025)051",
    journal = "JHEP",
    volume = "11",
    pages = "051",
    year = "2025"
}

@article{LHCb:2019tea,
    author = "Aaij, Roel and others",
    collaboration = "LHCb",
    title = "{Measurement of the $B_c^-$ meson production fraction and asymmetry in 7 and 13 TeV $pp$ collisions}",
    eprint = "1910.13404",
    archivePrefix = "arXiv",
    primaryClass = "hep-ex",
    reportNumber = "CERN-EP-2019-216, LHCb-PAPER-2019-033",
    doi = "10.1103/PhysRevD.100.112006",
    journal = "Phys. Rev. D",
    volume = "100",
    number = "11",
    pages = "112006",
    year = "2019"
}

@article{Fael:2022wfc,
    author = "Fael, Matteo and Rahimi, Muslem and Vos, K. Keri",
    title = "{New physics contributions to moments of inclusive b {\textrightarrow} c semileptonic decays}",
    eprint = "2208.04282",
    archivePrefix = "arXiv",
    primaryClass = "hep-ph",
    reportNumber = "P3H-22-090, SI-HEP-2022-21, TTP22-056, Nikhef 2022-010",
    doi = "10.1007/JHEP02(2023)086",
    journal = "JHEP",
    volume = "02",
    pages = "086",
    year = "2023"
}

@article{LHCb:2020cyw,
    author = "Aaij, Roel and others",
    collaboration = "LHCb",
    title = "{Measurement of $|V_{cb}|$ with $B_s^0 \to D_s^{(*)-} \mu^+ \nu_{\mu}$ decays}",
    eprint = "2001.03225",
    archivePrefix = "arXiv",
    primaryClass = "hep-ex",
    reportNumber = "LHCb-PAPER-2019-041, CERN-EP-2019-282",
    doi = "10.1103/PhysRevD.101.072004",
    journal = "Phys. Rev. D",
    volume = "101",
    number = "7",
    pages = "072004",
    year = "2020"
}

@article{LHCb:2020hpv,
    author = "Aaij, Roel and others",
    collaboration = "LHCb",
    title = "{Measurement of the shape of the $ {B}_s^0\to {D}_s^{\ast -}{\mu}^{+}{\nu}_{\mu } $ differential decay rate}",
    eprint = "2003.08453",
    archivePrefix = "arXiv",
    primaryClass = "hep-ex",
    reportNumber = "LHCb-PAPER-2019-046, CERN-EP-2020-026",
    doi = "10.1007/JHEP12(2020)144",
    journal = "JHEP",
    volume = "12",
    pages = "144",
    year = "2020"
}

@article{LHCb:2013vfg,
    author = "Aaij, R and others",
    collaboration = "LHCb",
    title = "{Measurement of the fragmentation fraction ratio $f_{s}/f_{d}$ and its dependence on $B$ meson kinematics}",
    eprint = "1301.5286",
    archivePrefix = "arXiv",
    primaryClass = "hep-ex",
    reportNumber = "CERN-PH-EP-2013-006, LHCB-PAPER-2012-037",
    doi = "10.1007/JHEP04(2013)001",
    journal = "JHEP",
    volume = "04",
    pages = "001",
    year = "2013"
}

@article{LHCb:2021qbv,
    author = "Aaij, Roel and others",
    collaboration = "LHCb",
    title = "{Precise measurement of the~$f_s/f_d$ ratio of fragmentation fractions and of $B^0_s$ decay branching fractions}",
    eprint = "2103.06810",
    archivePrefix = "arXiv",
    primaryClass = "hep-ex",
    reportNumber = "LHCb-PAPER-2020-046, CERN-EP-2021-027",
    doi = "10.1103/PhysRevD.104.032005",
    journal = "Phys. Rev. D",
    volume = "104",
    number = "3",
    pages = "032005",
    year = "2021"
}

@article{LHCb:2019lsv,
    author = "Aaij, Roel and others",
    collaboration = "LHCb",
    title = "{Measurement of $f_s / f_u$ Variation with Proton-Proton Collision Energy and $B$-Meson Kinematics}",
    eprint = "1910.09934",
    archivePrefix = "arXiv",
    primaryClass = "hep-ex",
    reportNumber = "CERN-EP-2019-209, LHCb-PAPER-2019-020",
    doi = "10.1103/PhysRevLett.124.122002",
    journal = "Phys. Rev. Lett.",
    volume = "124",
    number = "12",
    pages = "122002",
    year = "2020"
}

@article{LHCb:2020zae,
    author = "Aaij, R. and others",
    collaboration = "LHCb",
    title = "{Measurement of the branching fraction of the ${{B} ^0} {\rightarrow }{{D} ^+_{s}} {{\pi } ^-} $ decay}",
    eprint = "2010.11986",
    archivePrefix = "arXiv",
    primaryClass = "hep-ex",
    reportNumber = "LHCb-PAPER-2020-021, CERN-EP-2020-183",
    doi = "10.1140/epjc/s10052-020-08790-2",
    journal = "Eur. Phys. J. C",
    volume = "81",
    number = "4",
    pages = "314",
    year = "2021"
}

@article{LHCb:2011leg,
    author = "Aaij, R. and others",
    collaboration = "LHCb",
    title = "{Measurement of $b$-hadron production fractions in $7~\rm{TeV}$ pp collisions}",
    eprint = "1111.2357",
    archivePrefix = "arXiv",
    primaryClass = "hep-ex",
    reportNumber = "CERN-PH-EP-2011-172, LHCB-PAPER-2011-018",
    doi = "10.1103/PhysRevD.85.032008",
    journal = "Phys. Rev. D",
    volume = "85",
    pages = "032008",
    year = "2012"
}

@article{LHCb:2019fns,
    author = "Aaij, Roel and others",
    collaboration = "LHCb",
    title = "{Measurement of $b$ hadron fractions in 13 TeV $pp$ collisions}",
    eprint = "1902.06794",
    archivePrefix = "arXiv",
    primaryClass = "hep-ex",
    reportNumber = "CERN-EP-2019-016, LHCb-PAPER-2018-050",
    doi = "10.1103/PhysRevD.100.031102",
    journal = "Phys. Rev. D",
    volume = "100",
    number = "3",
    pages = "031102",
    year = "2019"
}

@article{Harrison:2025yan,
    author = "Harrison, Judd",
    collaboration = "HPQCD",
    title = "{Improved lattice QCD $B_c\to J/\psi$ vector, axial-vector, and tensor form factors}",
    eprint = "2503.15090",
    archivePrefix = "arXiv",
    primaryClass = "hep-lat",
    doi = "10.1103/wll6-z4cb",
    journal = "Phys. Rev. D",
    volume = "112",
    number = "3",
    pages = "034503",
    year = "2025"
}

@article{Harrison:2020gvo,
    author = "Harrison, Judd and Davies, Christine T. H. and Lytle, Andrew",
    collaboration = "HPQCD",
    title = "{$B_c \rightarrow J/\psi$ form factors for the full $q^2$ range from lattice QCD}",
    eprint = "2007.06957",
    archivePrefix = "arXiv",
    primaryClass = "hep-lat",
    doi = "10.1103/PhysRevD.102.094518",
    journal = "Phys. Rev. D",
    volume = "102",
    number = "9",
    pages = "094518",
    year = "2020"
}

@article{Iguro:2020cpg,
    author = "Iguro, Syuhei and Watanabe, Ryoutaro",
    title = "{Bayesian fit analysis to full distribution data of $ \overline{\mathrm{B}}\to {\mathrm{D}}^{\left(\ast \right)}\mathrm{\ell}\overline{\nu }:\left|{\mathrm{V}}_{\mathrm{cb}}\right| $ determination and new physics constraints}",
    eprint = "2004.10208",
    archivePrefix = "arXiv",
    primaryClass = "hep-ph",
    doi = "10.1007/JHEP08(2020)006",
    journal = "JHEP",
    volume = "08",
    number = "08",
    pages = "006",
    year = "2020"
}

@article{Jung:2018lfu,
    author = "Jung, Martin and Straub, David M.",
    title = "{Constraining new physics in $b\to c\ell\nu$ transitions}",
    eprint = "1801.01112",
    archivePrefix = "arXiv",
    primaryClass = "hep-ph",
    doi = "10.1007/JHEP01(2019)009",
    journal = "JHEP",
    volume = "01",
    pages = "009",
    year = "2019"
}

@article{Na:2015kha,
    author = "Na, Heechang and Bouchard, Chris M. and Lepage, G. Peter and Monahan, Chris and Shigemitsu, Junko",
    collaboration = "HPQCD",
    title = "{$B\to D \ell \nu$ form factors at nonzero recoil and extraction of $|V_{cb}|$}",
    eprint = "1505.03925",
    archivePrefix = "arXiv",
    primaryClass = "hep-lat",
    doi = "10.1103/PhysRevD.93.119906",
    journal = "Phys. Rev. D",
    volume = "92",
    number = "5",
    pages = "054510",
    year = "2015",
    note = "[Erratum: Phys.Rev.D 93, 119906 (2016)]"
}

@article{FermilabLattice:2015ilb,
    author = "Bailey, Jon A. and others",
    collaboration = "Fermilab Lattice, MILC",
    title = "{$B\to D\ell\nu$ form factors at nonzero recoil and $|V_{cb}|$ from 2+1-flavor lattice QCD}",
    eprint = "1503.07237",
    archivePrefix = "arXiv",
    primaryClass = "hep-lat",
    reportNumber = "FERMILAB-PUB-15-107-T",
    doi = "10.1103/PhysRevD.92.034506",
    journal = "Phys. Rev. D",
    volume = "92",
    number = "3",
    pages = "034506",
    year = "2015"
}

@article{Beneke:2021jhp,
    author = {Beneke, Martin and B{\"o}er, Philipp and Finauri, Gael and Vos, K. Keri},
    title = "{QED factorization of two-body non-leptonic and semi-leptonic B to charm decays}",
    eprint = "2107.03819",
    archivePrefix = "arXiv",
    primaryClass = "hep-ph",
    reportNumber = "TUM-HEP-1349/21, Nikhef-2021-015",
    doi = "10.1007/JHEP10(2021)223",
    journal = "JHEP",
    volume = "10",
    pages = "223",
    year = "2021"
}

@article{McLean:2019qcx,
    author = "McLean, E. and Davies, C. T. H. and Koponen, J. and Lytle, A. T.",
    title = "{$B_s\to D_s \ell\nu$ Form Factors for the full $q^2$ range from Lattice QCD with non-perturbatively normalized currents}",
    eprint = "1906.00701",
    archivePrefix = "arXiv",
    primaryClass = "hep-lat",
    doi = "10.1103/PhysRevD.101.074513",
    journal = "Phys. Rev. D",
    volume = "101",
    number = "7",
    pages = "074513",
    year = "2020"
}

@article{FermilabLattice:2021cdg,
    author = "Bazavov, A. and others",
    collaboration = "Fermilab Lattice, MILC",
    title = "{Semileptonic form factors for $B\rightarrow D^*\ell \nu $ at nonzero recoil from $2+1$-flavor lattice QCD: Fermilab Lattice~and~MILC~Collaborations}",
    eprint = "2105.14019",
    archivePrefix = "arXiv",
    primaryClass = "hep-lat",
    reportNumber = "FERMILAB-PUB-21-261-T~, FERMILAB-PUB-21/261-T",
    doi = "10.1140/epjc/s10052-022-10984-9",
    journal = "Eur. Phys. J. C",
    volume = "82",
    number = "12",
    pages = "1141",
    year = "2022",
    note = "[Erratum: Eur.Phys.J.C 83, 21 (2023)]"
}

@article{Aoki:2023qpa,
    author = "Aoki, Y. and Colquhoun, B. and Fukaya, H. and Hashimoto, S. and Kaneko, T. and Kellermann, R. and Koponen, J. and Kou, E.",
    collaboration = "JLQCD",
    title = {{$B\to D^*\ell\nu_\ell$ semileptonic form factors from lattice QCD with M{\"o}bius domain-wall quarks}},
    eprint = "2306.05657",
    archivePrefix = "arXiv",
    primaryClass = "hep-lat",
    reportNumber = "KEK-CP-393, OU-HET-1186",
    doi = "10.1103/PhysRevD.109.074503",
    journal = "Phys. Rev. D",
    volume = "109",
    number = "7",
    pages = "074503",
    year = "2024"
}

@article{Harrison:2023dzh,
    author = "Harrison, Judd and Davies, Christine T. H.",
    collaboration = "HPQCD",
    title = "{$B\to D^*$ and $B_s\to D_s^*$ vector, axial-vector and tensor form factors for the full $q^2$ range from lattice QCD}",
    eprint = "2304.03137",
    archivePrefix = "arXiv",
    primaryClass = "hep-lat",
    doi = "10.1103/PhysRevD.109.094515",
    journal = "Phys. Rev. D",
    volume = "109",
    number = "9",
    pages = "094515",
    year = "2024"
}

@article{Gubernari:2018wyi,
    author = "Gubernari, Nico and Kokulu, Ahmet and van Dyk, Danny",
    title = "{$B\to P$ and $B\to V$ Form Factors from $B$-Meson Light-Cone Sum Rules beyond Leading Twist}",
    eprint = "1811.00983",
    archivePrefix = "arXiv",
    primaryClass = "hep-ph",
    reportNumber = "EOS-2018-02, TUM-HEP-1172/18",
    doi = "10.1007/JHEP01(2019)150",
    journal = "JHEP",
    volume = "01",
    pages = "150",
    year = "2019"
}

@article{Bordone:2019guc,
    author = "Bordone, Marzia and Gubernari, Nico and van Dyk, Danny and Jung, Martin",
    title = "{Heavy-Quark expansion for ${{\bar{B}}_s\rightarrow D^{(*)}_s}$ form factors and unitarity bounds beyond the ${SU(3)_F}$ limit}",
    eprint = "1912.09335",
    archivePrefix = "arXiv",
    primaryClass = "hep-ph",
    reportNumber = "EOS-2019-04, P3H-19-050, SI-HEP-2019-20, TUM-HEP 1241/19",
    doi = "10.1140/epjc/s10052-020-7850-9",
    journal = "Eur. Phys. J. C",
    volume = "80",
    number = "4",
    pages = "347",
    year = "2020"
}

@article{Detmold:2015aaa,
    author = "Detmold, William and Lehner, Christoph and Meinel, Stefan",
    title = "{$\Lambda_b \to p \ell^- \bar{\nu}_\ell$ and $\Lambda_b \to \Lambda_c \ell^- \bar{\nu}_\ell$ form factors from lattice QCD with relativistic heavy quarks}",
    eprint = "1503.01421",
    archivePrefix = "arXiv",
    primaryClass = "hep-lat",
    reportNumber = "RBRC-1111",
    doi = "10.1103/PhysRevD.92.034503",
    journal = "Phys. Rev. D",
    volume = "92",
    number = "3",
    pages = "034503",
    year = "2015"
}

@article{Gambino:2023xoe,
    author = {Gambino, Paolo and Hashimoto, Shoji and M{\"a}chler, Sandro and Panero, Marco and Sanfilippo, Francesco and Simula, Silvano and Smecca, Antonio and Tantalo, Nazario},
    title = "{On the study of inclusive semileptonic decays of Bs-meson from lattice QCD{\ensuremath{\star}}}",
    eprint = "2311.09892",
    archivePrefix = "arXiv",
    primaryClass = "hep-lat",
    doi = "10.1393/ncc/i2024-24092-1",
    journal = "Nuovo Cim. C",
    volume = "47",
    number = "3",
    pages = "92",
    year = "2024"
}

@article{Finauri:2023kte,
    author = "Finauri, Gael and Gambino, Paolo",
    title = "{The q$^{2}$ moments in inclusive semileptonic B decays}",
    eprint = "2310.20324",
    archivePrefix = "arXiv",
    primaryClass = "hep-ph",
    reportNumber = "TUM-HEP 1477/23",
    doi = "10.1007/JHEP02(2024)206",
    journal = "JHEP",
    volume = "02",
    pages = "206",
    year = "2024"
}

@article{Belle:2006kgy,
    author = "Urquijo, P. and others",
    collaboration = "Belle",
    title = "{Moments of the electron energy spectrum and partial branching fraction of $B\to X_{(c)} e \nu$ decays at Belle}",
    eprint = "hep-ex/0610012",
    archivePrefix = "arXiv",
    reportNumber = "BELLE-2006-26, KEK-2006-37, BELLE-CONF-0667",
    doi = "10.1103/PhysRevD.75.032001",
    journal = "Phys. Rev. D",
    volume = "75",
    pages = "032001",
    year = "2007"
}

@article{BaBar:2006ztv,
    author = "Aubert, Bernard and others",
    collaboration = "BaBar",
    title = "{Measurement of the ratio $\mathcal{B}(B^+\to X e \nu)$ / $\mathcal{B}(B^0 \to X e \nu)$}",
    eprint = "hep-ex/0607111",
    archivePrefix = "arXiv",
    reportNumber = "SLAC-PUB-12027, BABAR-PUB-06-36, BABAR-PUB-06-036",
    doi = "10.1103/PhysRevD.74.091105",
    journal = "Phys. Rev. D",
    volume = "74",
    pages = "091105",
    year = "2006"
}

@article{CLEO:2004stg,
    author = "Mahmood, A. H. and others",
    collaboration = "CLEO",
    title = "{Measurement of the B-meson inclusive semileptonic branching fraction and electron energy moments}",
    eprint = "hep-ex/0403053",
    archivePrefix = "arXiv",
    reportNumber = "CLNS-04-1860, CLEO-04-02",
    doi = "10.1103/PhysRevD.70.032003",
    journal = "Phys. Rev. D",
    volume = "70",
    pages = "032003",
    year = "2004"
}

@article{LHCb:2021awg,
    author = "Aaij, Roel and others",
    collaboration = "LHCb",
    title = "{Measurement of the $B^0_s\to\mu^+\mu^-$ decay properties and search for the $B^0\to\mu^+\mu^-$ and $B^0_s\to\mu^+\mu^-\gamma$ decays}",
    eprint = "2108.09283",
    archivePrefix = "arXiv",
    primaryClass = "hep-ex",
    reportNumber = "CERN-EP-2021-133, LHCb-PAPER-2021-008",
    doi = "10.1103/PhysRevD.105.012010",
    journal = "Phys. Rev. D",
    volume = "105",
    number = "1",
    pages = "012010",
    year = "2022"
}

@article{CMS:2022mgd,
    author = "Tumasyan, Armen and others",
    collaboration = "CMS",
    title = "{Measurement of the $B^0_s\to\mu^+\mu^-$ decay properties and search for the $B^0\to\mu^+\mu^-$ decay in proton-proton collisions at $\sqrt{s}$ = 13 TeV}",
    eprint = "2212.10311",
    archivePrefix = "arXiv",
    primaryClass = "hep-ex",
    reportNumber = "CMS-BPH-21-006, CERN-EP-2022-270",
    doi = "10.1016/j.physletb.2023.137955",
    journal = "Phys. Lett. B",
    volume = "842",
    pages = "137955",
    year = "2023"
}

@article{Mannel:2023yqf,
    author = "Mannel, Thomas and Milutin, Ilija S. and Vos, K. Keri",
    title = "{Inclusive semileptonic $ b\to c\ell \overline{\nu} $ decays to order $ 1/{m}_b^5 $}",
    eprint = "2311.12002",
    archivePrefix = "arXiv",
    primaryClass = "hep-ph",
    reportNumber = "SI-HEP-2023-28, P3H-23-092, Nikhef-2023-04",
    doi = "10.1007/JHEP02(2024)226",
    journal = "JHEP",
    volume = "02",
    pages = "226",
    year = "2024",
    note = "[Erratum: JHEP 02, 167 (2025)]"
}

@article{Finauri:2025ost,
    author = "Finauri, Gael",
    title = "{Kinematic moments of $ \overline{B}\to {X}_c\ell {\overline{\nu}}_{\ell } $ to order $ \mathcal{O}\left({\Lambda}_{\textrm{QCD}}^5/{m}_b^5\right) $}",
    eprint = "2501.09090",
    archivePrefix = "arXiv",
    primaryClass = "hep-ph",
    doi = "10.1007/JHEP04(2025)112",
    journal = "JHEP",
    volume = "04",
    pages = "112",
    year = "2025"
}

@article{Bernlochner:2022ucr,
    author = "Bernlochner, Florian and Fael, Matteo and Olschewsky, Kevin and Persson, Eric and van Tonder, Raynette and Vos, K. Keri and Welsch, Maximilian",
    title = "{First extraction of inclusive V$_{cb}$ from q$^{2}$ moments}",
    eprint = "2205.10274",
    archivePrefix = "arXiv",
    primaryClass = "hep-ph",
    reportNumber = "Nikhef 2022-006, TTP22-031, P3H-22-052",
    doi = "10.1007/JHEP10(2022)068",
    journal = "JHEP",
    volume = "10",
    pages = "068",
    year = "2022"
}

@article{Herren:2026pni,
    author = "Herren, Florian and van Tonder, Raynette",
    title = "{Closing the knowledge gap in semileptonic $B\rightarrow X_c\ell\nu$ decays}",
    eprint = "2605.03021",
    archivePrefix = "arXiv",
    primaryClass = "hep-ph",
    reportNumber = "ZU-TH 17/26",
    month = "5",
    year = "2026"
}

@article{Gambino:2022dvu,
    author = {Gambino, Paolo and Hashimoto, Shoji and M{\"a}chler, Sandro and Panero, Marco and Sanfilippo, Francesco and Simula, Silvano and Smecca, Antonio and Tantalo, Nazario},
    title = "{Lattice QCD study of inclusive semileptonic decays of heavy mesons}",
    eprint = "2203.11762",
    archivePrefix = "arXiv",
    primaryClass = "hep-lat",
    doi = "10.1007/JHEP07(2022)083",
    journal = "JHEP",
    volume = "07",
    pages = "083",
    year = "2022"
}

@article{DeSantis:2025yfm,
    author = "De Santis, Alessandro and others",
    title = "{Inclusive Semileptonic Decays of the Ds Meson: Lattice QCD Confronts Experiments}",
    eprint = "2504.06064",
    archivePrefix = "arXiv",
    primaryClass = "hep-lat",
    reportNumber = "HIP-2025-10/TH",
    doi = "10.1103/snc6-cpz6",
    journal = "Phys. Rev. Lett.",
    volume = "135",
    number = "12",
    pages = "121901",
    year = "2025"
}

@article{DeSantis:2025qbb,
    author = "De Santis, Alessandro and others",
    title = "{Inclusive semileptonic decays of the Ds meson: A first-principles lattice QCD calculation}",
    eprint = "2504.06063",
    archivePrefix = "arXiv",
    primaryClass = "hep-lat",
    reportNumber = "HIP-2025-9/TH",
    doi = "10.1103/3cxg-k322",
    journal = "Phys. Rev. D",
    volume = "112",
    number = "5",
    pages = "054503",
    year = "2025"
}

@article{Carvunis:2025vab,
    author = {Carvunis, Alexandre and Finauri, Gael and Gambino, Paolo and Jung, Martin and M{\"a}chler, Sandro},
    title = "{New Physics in inclusive semileptonic B decays}",
    eprint = "2507.22123",
    archivePrefix = "arXiv",
    primaryClass = "hep-ph",
    doi = "10.1007/JHEP01(2026)037",
    journal = "JHEP",
    volume = "01",
    pages = "037",
    year = "2026"
}

@article{Mandal:2020htr,
    author = "Mandal, Rusa and Murgui, Clara and Pe{\~n}uelas, Ana and Pich, Antonio",
    title = "{The role of right-handed neutrinos in $b \to c \tau \bar{\nu}$ anomalies}",
    eprint = "2004.06726",
    archivePrefix = "arXiv",
    primaryClass = "hep-ph",
    reportNumber = "IFIC/20-14; FTUV/20-0414; SI-HEP-2020-XX, IFIC/20-14; FTUV/20-0414; SI-HEP-2020-10",
    doi = "10.1007/JHEP08(2020)022",
    journal = "JHEP",
    volume = "08",
    number = "08",
    pages = "022",
    year = "2020"
}

@article{Aebischer:2015fzz,
    author = "Aebischer, Jason and Crivellin, Andreas and Fael, Matteo and Greub, Christoph",
    title = "{Matching of gauge invariant dimension-six operators for $b\to s$ and $b\to c$ transitions}",
    eprint = "1512.02830",
    archivePrefix = "arXiv",
    primaryClass = "hep-ph",
    reportNumber = "CERN-PH-TH-2015-278",
    doi = "10.1007/JHEP05(2016)037",
    journal = "JHEP",
    volume = "05",
    pages = "037",
    year = "2016"
}

@article{Boyd:1997kz,
    author = "Boyd, C. Glenn and Grinstein, Benjamin and Lebed, Richard F.",
    title = "{Precision corrections to dispersive bounds on form-factors}",
    eprint = "hep-ph/9705252",
    archivePrefix = "arXiv",
    reportNumber = "CMU-HEP-97-07A, UCSD-PTH-97-12",
    doi = "10.1103/PhysRevD.56.6895",
    journal = "Phys. Rev. D",
    volume = "56",
    pages = "6895--6911",
    year = "1997"
}

@article{Gubernari:2023puw,
    author = "Gubernari, Nico and Reboud, M{\'e}ril and van Dyk, Danny and Virto, Javier",
    title = "{Dispersive analysis of $B \to K^{(*)}$ and $B_{s} \to \phi$ form factors}",
    eprint = "2305.06301",
    archivePrefix = "arXiv",
    primaryClass = "hep-ph",
    reportNumber = "EOS-2023-02, IPPP/23/22, P3H-23-026, SI-HEP-2023-09",
    doi = "10.1007/JHEP12(2023)153",
    journal = "JHEP",
    volume = "12",
    pages = "153",
    year = "2023",
    note = "[Erratum: JHEP 01, 125 (2025)]"
}

@article{EOSAuthors:2021xpv,
    author = "van Dyk, Danny and others",
    collaboration = "EOS Authors",
    title = "{\EOS: a software for flavor physics phenomenology}",
    eprint = "2111.15428",
    archivePrefix = "arXiv",
    primaryClass = "hep-ph",
    reportNumber = "EOS-2021-04, TUM-HEP 1371/21, P3H-21-094, SI-HEP-2021-32",
    doi = "10.1140/epjc/s10052-022-10177-4",
    journal = "Eur. Phys. J. C",
    volume = "82",
    number = "6",
    pages = "569",
    year = "2022"
}

@article{Becirevic:2012jf,
    author = "Be{\v{c}}irevi{\'c}, Damir and Ko{\v{s}}nik, Nejc and Tayduganov, Andrey",
    title = "{$\bar B\to D\tau\bar \nu_\tau$ vs. $\bar B\to D\mu\bar \nu_\mu$}",
    eprint = "1206.4977",
    archivePrefix = "arXiv",
    primaryClass = "hep-ph",
    reportNumber = "LPT-12-62, LAL-12-219",
    doi = "10.1016/j.physletb.2012.08.016",
    journal = "Phys. Lett. B",
    volume = "716",
    pages = "208--213",
    year = "2012"
}

@article{Duraisamy:2013pia,
    author = "Duraisamy, Murugeswaran and Datta, Alakabha",
    title = "{The Full $B \to D^{*} \tau^{-} \bar{\nu_\tau}$ Angular Distribution and CP violating Triple Products}",
    eprint = "1302.7031",
    archivePrefix = "arXiv",
    primaryClass = "hep-ph",
    doi = "10.1007/JHEP09(2013)059",
    journal = "JHEP",
    volume = "09",
    pages = "059",
    year = "2013"
}

@article{ParticleDataGroup:2024cfk,
    author = "Navas, S. and others",
    collaboration = "Particle Data Group",
    title = "{Review of particle physics}",
    doi = "10.1103/PhysRevD.110.030001",
    journal = "Phys. Rev. D",
    volume = "110",
    number = "3",
    pages = "030001",
    year = "2024"
}

@software{EOS:v1.0.20,
  author       = {Danny van Dyk and
                  Méril Reboud and
                  Matthew Kirk and
                  Nico Gubernari and
                  Philip Lüghausen and
                  Domagoj Leljak and
                  Stephan Kürten and
                  Ahmet Kokulu and
                  Carolina Bolognani and
                  Lorenz Gaertner and
                  Florian Herren and
                  Stefan Meiser and
                  Viktor Kuschke and
                  Filip Novak and
                  Christoph Bobeth and
                  Marta Burgos and
                  Ery McPartland and
                  Martin Ritter and
                  Romy O'Connor},
  title        = {{\EOS} version 1.0.20},
  month        = April,
  year         = 2026,
  publisher    = {Zenodo},
  version      = {v1.0.20},
  doi          = {\href{https://doi.org/10.5281/zenodo.19852912}{10.5281/zenodo.19852912}},
  url          = {https://doi.org/10.5281/zenodo.19852912}
}

@misc{EOS-DATA-2026-02,
  author       = "Bolognani, Carolina and Jung, Martin and Reboud, M\'eril and Vos, K. Keri",
  title        = "{\texttt{EOS/DATA-2026-02}: Supplementary material for \texttt{EOS/ANALYSIS-2024-03}}",
  month        = nov,
  year         = 2026,
  publisher    = {Zenodo},
  version      = {2026-02},
  doi          = {10.5281/zenodo.24168857},
  url          = {https://doi.org/10.5281/zenodo.24168857},
}

@article{Higson:2018,
	doi = {10.1007/s11222-018-9844-0},
	url = {https://doi.org/10.1007%2Fs11222-018-9844-0},
	year = 2018,
	month = {dec},
	publisher = {Springer Science and Business Media {LLC}},
	volume = {29},
	number = {5},
	pages = {891--913},
	author = {Edward Higson and Will Handley and Michael Hobson and Anthony Lasenby},
	title = {Dynamic nested sampling: an improved algorithm for parameter estimation and evidence calculation},
	journal = {Statistics and Computing}
}

@article{Flynn:2023qmi,
    author = {Flynn, J. M. and J{\"u}ttner, A. and Tsang, J. T.},
    title = "{Bayesian inference for form-factor fits regulated by unitarity and analyticity}",
    eprint = "2303.11285",
    archivePrefix = "arXiv",
    primaryClass = "hep-ph",
    reportNumber = "CERN-TH-2023-047",
    doi = "10.1007/JHEP12(2023)175",
    journal = "JHEP",
    volume = "12",
    pages = "175",
    year = "2023"
}

@article{Speagle:2020,
	doi = {10.1093/mnras/staa278},
	url = {https://doi.org/10.1093%2Fmnras%2Fstaa278},  
	year = 2020,
	month = {feb},
	publisher = {Oxford University Press ({OUP})},
	volume = {493},
	number = {3},
	pages = {3132--3158},
	author = {Joshua S Speagle},
	title = {\texttt{dynesty}: a dynamic nested sampling package for estimating {B}ayesian posteriors and evidences},
	journal = {Monthly Notices of the Royal Astronomical Society}
}

@misc{dynesty:v3.0.0,
	  author       = {Sergey Koposov and
		          Josh Speagle and
		          Kyle Barbary and
		          Gregory Ashton and
		          Ed Bennett and
		          Johannes Buchner and
		          Carl Scheffler and
		          Colm Talbot and
		          Ben Cook and
		          James Guillochon and
		          Patricio Cubillos and
		          Andrés Asensio Ramos and
		          Matthieu Dartiailh and
		          Ilya and
		          Erik Tollerud and
		          Dustin Lang and
		          Ben Johnson and
		          jtmendel and
		          Edward Higson and
		          Thomas Vandal and
		          Tansu Daylan and
		          Ruth Angus and
		          patelR and
		          Phillip Cargile and
		          Patrick Sheehan and
		          Matt Pitkin and
		          Matthew Kirk and
		          Lu Xu and
		          Joel Leja and
		          joezuntz},
	  title        = {joshspeagle/dynesty: v3.0.0},
	  month        = oct,
	  year         = 2025,
	  publisher    = {Zenodo},
	  version      = {v3.0.0},
	  doi          = {10.5281/zenodo.17268284},
	  url          = {https://doi.org/10.5281/zenodo.17268284},
}

@article{Rudolph:2018rzl,
    author = "Rudolph, Matthew",
    title = "{An experimentalist{\textquoteright}s guide to the semileptonic bottom to charm branching fractions}",
    eprint = "1805.05659",
    archivePrefix = "arXiv",
    primaryClass = "hep-ph",
    doi = "10.1142/S0217751X18501762",
    journal = "Int. J. Mod. Phys. A",
    volume = "33",
    number = "32",
    pages = "1850176",
    year = "2018"
}

@article{Bernlochner:2012bc,
    author = "Bernlochner, Florian U. and Ligeti, Zoltan and Turczyk, Sascha",
    title = "{A Proposal to solve some puzzles in semileptonic B decays}",
    eprint = "1202.1834",
    archivePrefix = "arXiv",
    primaryClass = "hep-ph",
    doi = "10.1103/PhysRevD.85.094033",
    journal = "Phys. Rev. D",
    volume = "85",
    pages = "094033",
    year = "2012"
}

@article{Herren:2026pbh,
    author = "Herren, Florian and van Tonder, Raynette",
    title = "{On the simulated kinematic distributions of semileptonic $B$ decays}",
    eprint = "2602.18378",
    archivePrefix = "arXiv",
    primaryClass = "hep-ph",
    reportNumber = "ZU-TH 08/26",
    month = "2",
    year = "2026"
}

@article{Jung:2026ewj,
    author = "Jung, Martin and Schacht, Stefan",
    title = "{$\bar B\to D^{(*)}\ell\bar \nu$ Branching Ratios and Evidence for Isospin Breaking in $\Upsilon(4S)$ Decays}",
    eprint = "2604.08391",
    archivePrefix = "arXiv",
    primaryClass = "hep-ph",
    reportNumber = "IPPP/26/30",
    month = "4",
    year = "2026"
}

@article{Bernlochner:2023bad,
    author = "Bernlochner, Florian and Jung, Martin and Khan, Munira and Landsberg, Greg and Ligeti, Zoltan",
    title = "{Novel approaches to determine $B^\pm$ and $B^0$ meson production fractions}",
    eprint = "2306.04686",
    archivePrefix = "arXiv",
    primaryClass = "hep-ph",
    doi = "10.1103/PhysRevD.110.014007",
    journal = "Phys. Rev. D",
    volume = "110",
    number = "1",
    pages = "014007",
    year = "2024"
}

@article{Belle:2012mwf,
    author = "Oswald, C. and others",
    collaboration = "Belle",
    title = "{Measurement of the inclusive semileptonic branching fraction $\mathcal{B}(B_s^0 \to X^- \ell^+ \nu_{\ell})$ at Belle}",
    eprint = "1212.6400",
    archivePrefix = "arXiv",
    primaryClass = "hep-ex",
    reportNumber = "BELLE-PREPRINT-2012-29, KEK-PREPRINT-2012-34",
    doi = "10.1103/PhysRevD.87.072008",
    journal = "Phys. Rev. D",
    volume = "87",
    number = "7",
    pages = "072008",
    year = "2013",
    note = "[Erratum: Phys.Rev.D 90, 119901 (2014)]"
}

@article{Belle:2015ftp,
    author = "Oswald, C. and others",
    collaboration = "Belle",
    title = "{Semi-inclusive studies of semileptonic $B_s$ decays at Belle}",
    eprint = "1504.02004",
    archivePrefix = "arXiv",
    primaryClass = "hep-ex",
    reportNumber = "BELLE-PREPRINT-2015-4, KEK-PREPRINT-2015-1",
    doi = "10.1103/PhysRevD.92.072013",
    journal = "Phys. Rev. D",
    volume = "92",
    number = "7",
    pages = "072013",
    year = "2015"
}

@article{Dowdall:2012ab,
    author = "Dowdall, R. J. and Davies, C. T. H. and Hammant, T. C. and Horgan, R. R.",
    title = "{Precise heavy-light meson masses and hyperfine splittings from lattice QCD including charm quarks in the sea}",
    eprint = "1207.5149",
    archivePrefix = "arXiv",
    primaryClass = "hep-lat",
    doi = "10.1103/PhysRevD.86.094510",
    journal = "Phys. Rev. D",
    volume = "86",
    pages = "094510",
    year = "2012"
}

@article{Godfrey:2004ya,
    author = "Godfrey, Stephen",
    title = "{Spectroscopy of $B_c$ mesons in the relativized quark model}",
    eprint = "hep-ph/0406228",
    archivePrefix = "arXiv",
    reportNumber = "ADP-04-17-T599",
    doi = "10.1103/PhysRevD.70.054017",
    journal = "Phys. Rev. D",
    volume = "70",
    pages = "054017",
    year = "2004"
}

@article{CMS:2019uhm,
    author = "Sirunyan, Albert M and others",
    collaboration = "CMS",
    title = "{Observation of Two Excited B$^+_\mathrm{c}$ States and Measurement of the B$^+_\mathrm{c}$(2S) Mass in pp Collisions at $\sqrt{s} =$ 13 TeV}",
    eprint = "1902.00571",
    archivePrefix = "arXiv",
    primaryClass = "hep-ex",
    reportNumber = "CMS-BPH-18-007, CERN-EP-2019-014",
    doi = "10.1103/PhysRevLett.122.132001",
    journal = "Phys. Rev. Lett.",
    volume = "122",
    number = "13",
    pages = "132001",
    year = "2019"
}

@article{LHCb:2019bem,
    author = "Aaij, Roel and others",
    collaboration = "LHCb",
    title = "{Observation of an excited $B_c^+$ state}",
    eprint = "1904.00081",
    archivePrefix = "arXiv",
    primaryClass = "hep-ex",
    reportNumber = "CERN-EP-2019-050, LHCb-PAPER-2019-007",
    doi = "10.1103/PhysRevLett.122.232001",
    journal = "Phys. Rev. Lett.",
    volume = "122",
    number = "23",
    pages = "232001",
    year = "2019"
}

@article{LHCb:2022piu,
    author = "Aaij, R. and others",
    collaboration = "LHCb",
    title = "{Observation of the decay $ \Lambda_b^0\rightarrow \Lambda_c^+\tau^-\overline{\nu}_{\tau}$}",
    eprint = "2201.03497",
    archivePrefix = "arXiv",
    primaryClass = "hep-ex",
    reportNumber = "LHCb-PAPER-2021-044, CERN-EP-2021-265",
    doi = "10.1103/PhysRevLett.128.191803",
    journal = "Phys. Rev. Lett.",
    volume = "128",
    number = "19",
    pages = "191803",
    year = "2022"
}

@article{LHCb:2015eia,
    author = "Aaij, Roel and others",
    collaboration = "LHCb",
    title = "{Determination of the quark coupling strength $|V_{ub}|$ using baryonic decays}",
    eprint = "1504.01568",
    archivePrefix = "arXiv",
    primaryClass = "hep-ex",
    reportNumber = "CERN-PH-EP-2015-084, LHCB-PAPER-2015-013",
    doi = "10.1038/nphys3415",
    journal = "Nature Phys.",
    volume = "11",
    pages = "743--747",
    year = "2015"
}

@article{HeavyFlavorAveragingGroupHFLAV:2024ctg,
    author = "Banerjee, Sw. and others",
    collaboration = "Heavy Flavor Averaging Group (HFLAV)",
    title = "{Averages of b-hadron, c-hadron, and {\ensuremath{\tau}}-lepton properties as of 2023}",
    eprint = "2411.18639",
    archivePrefix = "arXiv",
    primaryClass = "hep-ex",
    doi = "10.1103/x87q-tld5",
    journal = "Phys. Rev. D",
    volume = "113",
    number = "1",
    pages = "012008",
    year = "2026"
}

@article{Amhis:2021cfy,
    author = "Amhis, Yasmine and Hartmann, Marie and Helsens, Cl{\'e}ment and Hill, Donal and Sumensari, Olcyr",
    title = "{Prospects for $B_c^+\to\tau^+\nu_\tau$ at FCC-ee}",
    eprint = "2105.13330",
    archivePrefix = "arXiv",
    primaryClass = "hep-ex",
    doi = "10.1007/JHEP12(2021)133",
    journal = "JHEP",
    volume = "12",
    pages = "133",
    year = "2021"
}

@article{Meiser:2024zea,
    author = "Meiser, Stefan and van Dyk, Danny and Virto, Javier",
    title = "{Towards a global analysis of the b {\textrightarrow}$ c\overline{u}q $ puzzle}",
    eprint = "2411.09458",
    archivePrefix = "arXiv",
    primaryClass = "hep-ph",
    reportNumber = "EOS-2024-04, IPPP/24/71",
    doi = "10.1007/JHEP06(2025)019",
    journal = "JHEP",
    volume = "06",
    pages = "019",
    year = "2025"
}

\end{document}